\documentclass[12pt, preprint]{aastex}

\usepackage{pdfsync}
\usepackage{amsmath}
\usepackage{natbib}

\DeclareMathSymbol{\varOmega}{\mathord}{letters}{"0A}
\DeclareMathSymbol{\varSigma}{\mathord}{letters}{"06}
\DeclareMathSymbol{\varPsi}{\mathord}{letters}{"09}

\def \kagara {=\hskip -5.2 pt $|$\hskip 2.5 pt K\c{\'a}g\'ara}

\def \RH {R_\mathrm{H}}

\begin{document}


\shorttitle{Planetesimal Formation}
\shortauthors{Nesvorny et al.}

\title{Binary Planetesimal Formation from Gravitationally \\Collapsing Pebble Clouds}
\author{David Nesvorn\'y$^1$,  Rixin Li$^2$, Jacob B. Simon$^3$, Andrew N.\ Youdin$^{2,4}$, Derek C. Richardson$^5$, Raphael Marschall$^1$, 
William M. Grundy$^6$}
\affil{(1) Department of Space Studies, Southwest Research Institute, 1050 Walnut St., Suite 300, Boulder, CO 80302, USA}
\affil{(2) Steward Observatory \& Department of Astronomy, University of Arizona, 933 N. Cherry Avenue, Tucson, AZ, USA}
\affil{(3) Physics and Astronomy Department, Iowa State University, 2323 Osborn Dr., Ames, IA 50011-1026, USA}
\affil{(4) Lunar and Planetary Laboratory, University of Arizona, 1629 E University Blvd, Tucson, AZ 85721, USA}
\affil{(5) Department of Astronomy, University of Maryland, College Park, MD 20742-2421, USA}
\affil{(6) Lowell Observatory, 1400 W. Mars Hill Rd., Flagstaff, AZ 86001, USA}

\begin{abstract}
Planetesimals are compact astrophysical objects roughly 1-1000 km in size, massive enough to be held together by gravity. They can grow 
by accreting material to become full-size planets. Planetesimals themselves are thought to form by complex physical processes from small 
grains in protoplanetary disks. The streaming instability (SI) model states that mm/cm-size particles (pebbles) are aerodynamically 
collected into self-gravitating clouds which then directly collapse into planetesimals. Here we analyze {\tt ATHENA} simulations of 
the SI to characterize the initial properties (e.g., rotation) of pebble clouds. Their gravitational collapse is followed with the 
{\tt PKDGRAV} $N$-body code, which has been modified to realistically account for pebble collisions.
We find that pebble clouds rapidly collapse into short-lived disk structures from which planetesimals form. The planetesimal properties 
depend on the cloud's scaled angular momentum, $l=L/(MR_{\rm H}^2\Omega)$, where $L$ and $M$ are the angular momentum and mass, $R_{\rm H}$ 
is the Hill radius, and $\Omega$ is the orbital frequency. Low-$l$ pebble clouds produce tight (or contact) binaries and single planetesimals. 
Compact high-$l$ clouds give birth to binary planetesimals with attributes that closely resemble the equal-size binaries found in the 
Kuiper belt. Significantly, the SI-triggered gravitational collapse can explain the angular momentum distribution of known equal-size 
binaries -- a result pending verification from studies with improved resolution. About 10\% of collapse simulations produce hierarchical 
systems with two or more large moons. These systems should be found in the Kuiper belt when observations reach the threshold sensitivity. 
 
\end{abstract}

\section{Introduction}

Planet formation is a fundamental scientific problem in astrophysics. We want to know how the solar system planets formed, how planets populate the
universe, and how they become habitable and can potentially host extraterrestrial life. An important stage of planet formation
is the formation of ``small planets'', or planetesimals, that are 1-1000 km in size. There are three populations of leftover planetesimals 
in the solar system: the asteroid and Kuiper belts, and the Oort cloud. The size distributions of asteroids and Kuiper belt objects (KBOs) 
provide important constraints on planetesimal formation. In both cases there is a marked excess of 100-km class bodies,
which is thought to reflect the preferred size of objects that are assembled by planetesimal formation processes 
(e.g. Goldreich \& Ward 1973, Sekiya 1983, Youdin \& Shu 2002, Morbidelli et al. 2009). 
This signature, however, is entangled with the collisional evolution of asteroids/KBOs which has altered their initial distributions 
to some degree (e.g., Pan \& Sari 2005, Bottke et al.\ 2005a,b, Fraser 2009, Weidenschilling 2011, Nesvorn\'y \& Vokrouhlick\'y 2019). 

Asteroid collisions can liberate small fragments, known as meteoroids, that dynamically evolve to occasionally reach the planet-crossing 
orbits. They can impact on the Earth and, if large enough to survive atmospheric ablation, may end up being found on the ground, 
classified in our meteorite collection and studied. The isotopic chronology of meteorites suggests a relatively narrow range of the asteroid 
formation ages, $\sim0.5$-3 Myr for iron meteorites and $\sim1$-5 Myr for chondritic/achondritic meteorites (ages referred to CAI formation 
4567.3 Gyr ago, Connelly et al.\ 2012; see Kleine et al.\ 2009, 2020 for reviews). These results imply that asteroids most likely formed 
during the lifetime of the solar gas nebula, if protoplanetary disk ages can be taken as a guide, because gas disks typically last 
$\sim2$-10~Myr (Haisch et al.\ 2001, Williams \& Cieza 2011).

Observations of protoplanetary and debris disks provide key constraints on planetesimal formation. Dusty/icy particles 
detected in debris disks are produced by the collisional grinding of planetesimal belts. 
The ubiquity of the debris disk phenomenon suggests that planetesimal belts are common thus highlighting the
significance of planetesimals for planet formation (Wyatt 2008). ALMA observations of protoplanetary disks reveal small-scale 
structures such as rings, spirals and arcs (Andrews 2020). At least some of these structures may be tracing gas pressure maxima 
where particles pile up. These would be the prime locations for planetesimal formation because streaming and other instabilities 
--thought to be responsible for the formation of planetesimals (Sect.\ 3)-- efficiently operate in protoplanetary disks if the 
dust-to-gas ratio is enhanced to some degree (Carrera et al.\ 2020).

Closer to home, HST observations of KBOs reveal that these objects are often found in wide binaries with nearly equal-size components. 
The equal-size binaries are particularly prominent in a dynamical class of KBOs known as Cold Classicals (CCs;  heliocentric orbits with 
semimajor axes $a=42$-47 au, eccentricities $e<0.15$ and inclinations $i<5^\circ$), where the binary fraction is estimated to be $>30$\% 
(Noll et al.\ 2008, Fraser et al.\ 2017). The equal-size binaries cannot form in the present Kuiper belt. Instead, they must have formed 
during the formation of KBOs themselves or by early capture (e.g., Goldreich et al.\ 2002, Nesvorn\'y et al.\ 2010). The KBO binaries 
thus represent an important clue to planetesimal formation. The matching colors of binary components (Benecchi et al.\ 2011) imply that 
each binary formed with a uniform compositional mix, as expected for gravitational collapse (but not random capture).

\section{Properties of known KBO binaries}
 
\begin{figure}
\epsscale{0.7}
\plotone{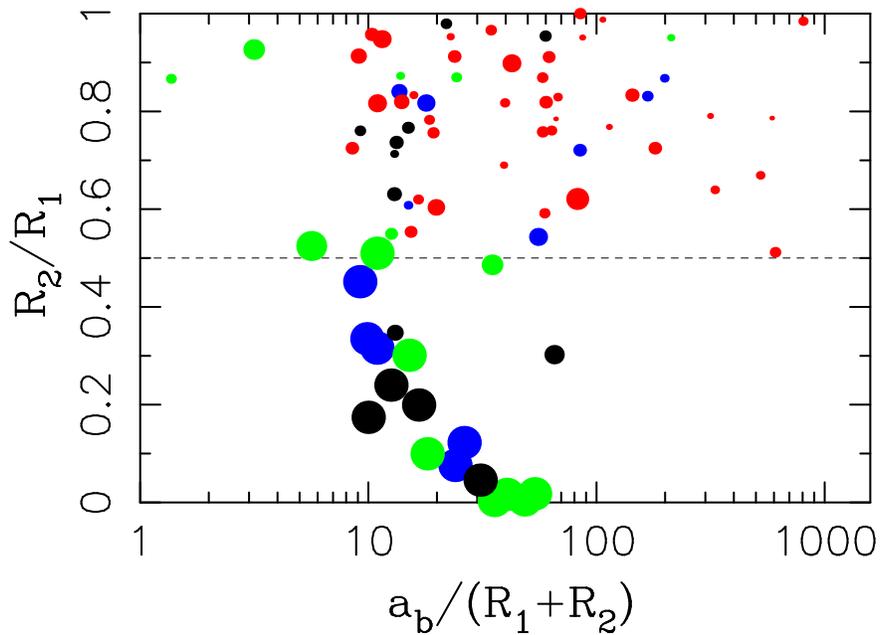}
\caption{Properties of known KBO binaries/satellites. The color code indicates the membership of binaries in different 
dynamical classes (red for CCs, green for Plutinos, blue for Hot Classicals, black for everything else). The symbol size 
linearly correlates with the primary radius for $R_1 \leq 250$ km and is fixed for $R_1>250$ km. For example, Pluto's 
four small satellites are shown by large green dots at the bottom 
of the plot and Charon is the largest green dot near the $R_2/R_1=0.5$ line. The unequal-size systems ($R_2/R_1\lesssim0.5$) 
are detected around large primaries in the dynamically hot populations. Most known equal-size binaries ($R_2/R_1>0.5$) 
are in the CC population (shown in red in the upper half of the plot).} 
\label{realbin}
\end{figure}

A catalog of physical and orbital properties of binary bodies is maintained by W. R. Johnston (Johnston 2018) on the NASA Planetary Data 
System (PDS) node.\footnote{{\tt https://sbn.psi.edu/pds/resource/binmp.html}} Figure \ref{realbin} shows the basic properties of 
known KBO binaries/satellites. Two notable features are apparent in the plot. First, the unequal-size binaries with a large primary 
and a small moon ($R_2/R_1<0.5$, where $R_1$ and $R_2$ denote the primary and secondary radii) are mainly detected 
in the hot population (Plutinos in the 3:2 resonance with Neptune and other resonant populations, Hot Classicals, scattered disk, 
etc.).\footnote{See Gladman et al.\ (2008) for a definition of different dynamical categories.} These moons are thought to have accreted 
around their primaries from impact-generated disks (Canup 2005, Leinhardt et al.\ 2010). They are absent in the CC population either because 
they did not form or because bodies in the cold population are generally smaller and the moons with $R_2<0.5\,R_1$ around small primaries 
are difficult to detect. Second, most known equal-size binaries with $R_2/R_1>0.5$ appear in the CC population (40 out of 65 known; 
shown in red in Fig. \ref{realbin}; Table 1). 

The most straightforward interpretation of these differences is that collisions played an important role in shaping the hot population, 
whereas the collisional evolution of CCs was relatively modest (e.g., Nesvorn\'y et al.\ 2011, Parker \& Kavelaars 2012). 
The collisional activity in the present Kuiper belt is low and not very different between the hot and cold populations. This means that 
bodies in the hot population must have collisionally evolved {\it before} they were implanted into the Kuiper belt, 
probably during an early stage when they were embedded in the massive planetesimal disk (Levison et al.\ 2008). The CC population 
did not follow the same evolution path most likely because the outer extension of the planetesimal disk at $>40$~au had a relatively 
low mass (Morbidelli \& Nesvorn\'y 2020). In addition, it has been pointed out (Parker \& Kavelaars 2010) that the wide binaries in 
the CC population ($a_{\rm b}/R_{\rm b}>100$, where $a_{\rm b}$ is the semimajor axis of binary orbit and $R_{\rm b}=R_1+R_2$) would not 
dynamically survive during the implantation process, thus strengthening the idea that most of them formed beyond 40~au.\footnote{The 
photometrically {\it blue} binaries, which represent $\sim$25\% of equal-size binaries detected in the CC population, probably formed 
at 30-40 au and were `pushed out' during Neptune's migration (Fraser et al.\ 2017).} If so, they largely avoided close encounters 
with Neptune (see Shannon \& Dawson 2018 and Nesvorn\'y \& Vokrouhlick\'y 2019 for the dynamical effects affecting KBO binaries).
In summary, given their distant low-inclination orbits, low total mass and large binary fraction, {\it CCs are thought to be the 
least dynamically/collisionally evolved population of planetesimals in the solar system}. 

\section{Planetesimal formation}

During the earliest stages of planet formation, small grains condense in a protoplanetary nebula and grow to larger ice/dust aggregates 
by sticking to each other. The growth stalls near cm sizes, because molecular forces are not strong enough to hold large particles
together, and because large particles spiral toward the central star before they can grow (e.g., Birnstiel et al.\ 2016). An aerodynamic 
interaction between particles and nebular gas is important during these stages. The interaction gives rise to the streaming instability 
(SI; Youdin \& Goodman 2005) which is thought to be particularly important for planetesimal formation. In the limit of low pebble-gas 
ratios, the SI is one of a broad class of resonant drag instabilities (Squire \& Hopkins 2018). The SI occurs when an initially small 
over-density of pebbles accelerates the gas. This perturbation launches a wave that amplifies 
pebble density as it oscillates. Strong pebble clumping eventually triggers gravitational collapse into planetesimals (e.g., Youdin \& 
Johansen 2007, Johansen \& Youdin 2007, Johansen et al.\ 2009, Simon et al.\ 2017, Li et al.\ 2018).

The initial stages of particle concentrations by the SI can be studied with specialized hydrocodes (e.g., {\tt ATHENA} with the particle 
module of Bai \& Stone 2010). Modern hydrocode simulations of the SI, however, do not have an adequate spatial resolution to follow 
the gravitational collapse to completion. Moreover, once the pebble density within a collapsing cloud exceeds the gas density, aerodynamic 
effects of gas cease to be important, and detailed hydrodynamic calculations are no longer required (Nesvorn\'y, Youdin \& Richardson 2010; 
hereafter NYR10; Appendix A). Instead, one has to realistically model pebble collisions that damp random speeds and stimulate growth 
(Wahlberg Jansson \& Johansen 2014). 

In our previous work (NYR10), we studied gravitational collapse with a cosmological $N$-body code known as {\tt PKDGRAV} (Stadel 2001). {\tt 
PKDGRAV} is a scalable, parallel tree code that is the fastest code available to us for these type of calculations. A unique feature of 
{\tt PKDGRAV} is the ability to rapidly detect and realistically treat collisions between particles (Richardson et al.\ 2000). 
We showed that (i) the gravitational collapse model is capable of producing a very large binary fraction consistent with observations, and (ii) 
planetesimal binaries that form in gravitational collapse have nearly equal-size components and large separations, just as needed to 
explain observations (also see Robinson et al.\ 2020). 

NYR10 adopted idealized initial conditions for gravitational collapse -- rigidly rotating spherical clouds of particles with uniform 
spatial density. In the present work, in a step toward realism, we use the results of {\tt ATHENA} SI simulations to set up the initial 
conditions for {\tt PKDGRAV}. This turns out to be important because the results are sensitive not only to the cloud's global rotation 
but also to its particular radial profile. The firm connection of the present investigation to the SI simulations allows us, for the 
first time, to formulate expectations of the SI model for the formation of planetesimal binaries.

\section{Superparticle method}

In NYR10, individual {\tt PKDGRAV} particles were artificially inflated to mimic a very large collisional cross section of pebbles. We simply 
scaled up the radius of {\tt PKDGRAV} particles by a multiplication factor that was the same for all particles and unchanging with time. 
This is not ideal for several different reasons. Critically, the method with a fixed inflation factor does not properly account for the 
kinetic energy loss during inelastic collisions. 

A more realistic treatment of pebble collisions is now available (Nesvorn\'y et al.\ 2020). Consider two particle systems, one that consists 
of real particles (RPs; i.e., pebbles) and another one that consists of simulated superparticles (SPs) in {\tt PKDGRAV}. The number of SPs 
is vastly smaller than the number of RPs. Inelastic collisions happen in both systems; they can result in inelastic bounces or mergers of 
particles, and the loss of kinetic energy. The method described in Nesvorn\'y et al.\ (2020) shows how to best deal with collisions in the SP 
system such that it statistically reproduces the behavior of the RP system (i.e., overall energy loss, particle growth). In a special case, 
when colliding SPs $i$ and $j$ represent the same number of RPs, $n_i=n_j$ ($=n$), the effective SP radii are set to $R_i = \sqrt{n} r_i$ 
and $R_j = \sqrt{n} r_j$, where $r_i$ and $r_j$ are the RP radii. This guarantees that both the energy loss and growth rate are the same 
in SP and RP systems. 

If $n_i \neq n_j$, the energy loss and growth are not both exactly replicated, but approximate solutions exist. Specifically, we 
implemented an algorithm in {\tt PKDGRAV} that exactly replicates the energy loss in the RP system. See Nesvorn\'y et al.\ (2020) for a 
complete description of the method. If SPs merge, this represents a merger 
of some large number of RPs.  If SPs $i$ and $j$ with masses $m_i$ and $m_j$ merge to a new SP $k$, the masses simply add $m_k = m_i + m_j$.  
However, there is some choice as to how the underlying true particle masses combine. For example, if SPs $i$ and $j$ represent RPs with masses 
$m_i'$ and $m_j'$, we can assume that the new SP will represent RPs with mass $m_k' = m_i' + m_j'$. The new sampling rate is $n_k = m_k/m_k'$. 
The sampling rate is reduced to $n_{k}' =  \max[n_{k} (1 - f m_j/m_k), 1]$ for $m_i > m_j$ and constant $0<f<2$, such that SPs gradually 
approach RPs.

We tested the method described above and found that $f \sim 1$ leads to a growth rate that reasonably well replicates the growth rate 
in the RP system. The tests were performed against a reference case with $10^6$ RPs (Nesvorn\'y et al.\ 2020). Here we deal with systems 
where each cloud should consist of $\sim 10^{20}$ pebbles. Each SP must therefore initially represent a much larger number of RPs and 
$f>1$ is needed to more aggressively reduce the sampling rate (and converge toward $n=1$ at the end of the simulations; recall that 
the sampling rate is the superparticle-to-pebble mass ratio, or equivalently, the number of pebbles each superparticle must represent). 
Here we therefore study cases with $1 \leq f < 2$ (note that $f<2$ by definition such that the sampling rate remains positive for
collisions of equal-mass SPs) to understand the dependence of the results on this parameter. In essence, this procedure reflects 
our ignorance about the behavior of the RP system on unresolved, small size scales. 

Another important parameter is the coefficient of restitution, $C_{\rm R}$, which defines how much kinetic energy is lost in inelastic 
collisions. The coefficient of restitution depends on the RP size (pebble/planetesimal), material properties (rock/ice) and 
structure (compact particle/particle aggregate) and collision velocities. For example, Hill et al.\ (2015) conducted experiments 
with solid ice spheres and fragments, 5-10 millimeters in size, and collision velocities $v_{\rm col} = 0.27$ and 0.51 m s$^{-1}$. They 
found a generous range of $C_{\rm R}$ values between 0.08 and 0.65. These results are not directly applicable to  
particle aggregates that may have grown by sticking of $\sim \mu$m grains (see Blum et al.\ 2018 for a recent review).    

Given these uncertainties here we adopt a simple configuration. The tangential component of $C_{\rm R}$ is set to 1
(i.e, no energy loss) to avoid coupling between particle rotation and translation during grazing collisions. A study of this 
effect on planetesimal rotation is left for future work. The normal component of $C_{\rm R}$ is set to a fixed value. Note that this means 
that pebble collisions are treated in much the same way as planetesimal collisions, as if one single value of `effective' 
$C_{\rm R}$ could capture the statistical behavior of the system. Specifically, we test cases with $C_{\rm R}=0.1$, 0.3, 
0.5, 0.7 and 0.9. Particle fragmentation is not considered in this work because the collision velocities between pebbles are generally 
low ($v_{\rm col} < 1$ m s$^{-1}$; see Wahlberg Jansson et al.\ 2017). 

\section{{\tt ATHENA} simulations of the Streaming Instability}

We use a suite of vertically stratified 3D simulations of the SI (Simon et al.\ 2017; Li et al.\ 2018, 2018; Nesvorn\'y et al.\ 2019) 
to set up the initial conditions for the collapse simulations. The SI simulations were performed with the {\tt ATHENA} code (Stone et al.\ 2008), 
which accounts for the hydrodynamic flow of gas, aerodynamic forces on particles, backreaction of particles on the gas flow, and particle 
self-gravity. The simulations had a relatively high resolution with at least 512$^3$ grid 
cells and more than $1.3\times10^8$ particles. Each simulation 
was parametrized by the dimensionless stopping time, $\tau=t_{\rm stop} \Omega$, where $\Omega$ is the Keplerian frequency, and the local 
particle-to-gas column density ratio, $Z$. Here we analyze runs A12, B22 and C203 with $\tau=0.3$-2, which would correspond to mm/cm-size 
pebbles at 45 au (at least for some reasonable disk models), and $Z=0.02$-0.1 (Table~2). See Nesvorn\'y et al.\ 
(2019) for a detailed description of the simulation parameters. 

\begin{figure}
\epsscale{1.0}
\plotone{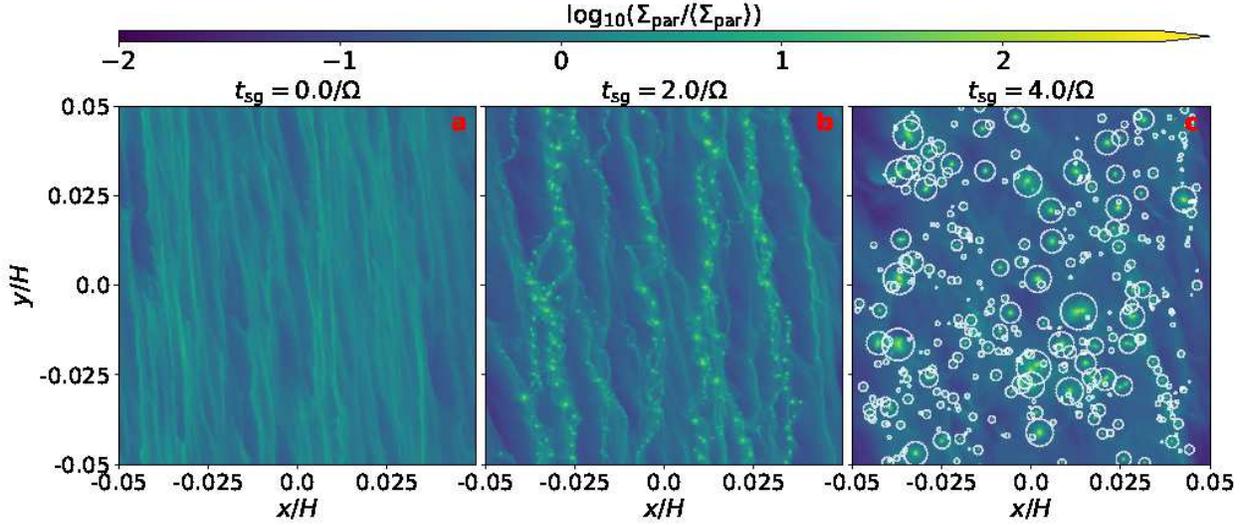} 
\caption{Three snapshots from a 3D {\tt ATHENA} simulation of the SI where the nonlinear particle clumping triggers gravitational 
collapse into planetesimals. The plots show the vertically-integrated density of solids ($\Sigma_{\rm par}$), projected on the disk plane, 
relative to the initially uniform surface density ($\langle\Sigma_{\rm par}\rangle$). The $x$ and $y$ coordinates show the shearing box dimensions 
in units of the gas disk scale height, $H_{\rm g}$ (the Sun is to the left, the orbital velocity vector points up). Time $t$ increases 
from left to right as labeled ($t_{\rm sg} = t - t_0$; particle self-gravity was switched on at $t_0=36/\Omega$). Azimuthal filaments have 
already formed at $t_{\rm sg}=0$ (panel a). In b and c, the filaments fragment into gravitationally bound clumps. The circles in panel 
c depict the Hill spheres of clumps that were identified by the {\tt PLAN} algorithm. Reproduced from the A12 run described in 
Nesvorn\'y et al. (2019).}
\label{athena}
\end{figure}

The A12 and B22 simulations covered a time interval of $\simeq 50/\Omega$, corresponding to $\simeq$ 8 orbital periods (or $\simeq2400$ 
yr at 45 au). The C203 simulation was run up to time $t\simeq130/\Omega$. The gravity in {\tt ATHENA} was switched on at $t_0=36/\Omega$, 
$37/\Omega$ and $110/\Omega$ in A12, B22 and C203, respectively. As the time progresses in the simulations, dense azimuthal filaments form, 
fragment and eventually, after $t_0$, condense into hundreds of gravitationally-bound clumps (Fig. \ref{athena}). We used an efficient 
tree-based code known as the PLanetesimal ANalyzer, or {\tt PLAN} for short (Li 2019, Li et al. 2019), to identify all clumps 
(Fig. \ref{athena}c).\footnote{{\tt PLAN} takes {\tt ATHENA}'s particle data and organizes them into a Barnes-Hut tree (Barnes \& Hut 
1986), which enables the code to quickly group particles into dense clumps based on the method described in Eisenstein \& Hut (1998). 
The clumps are then examined. They can be merged (if bound to each other) or discarded (if their Hill radius is smaller than a grid 
cell). The final product of {\tt PLAN} is a list of individual clumps that are held together by gravity (see Li et al. 2019 for 
more details).} For each clump we measure the total angular momentum, $L$, and its $z$-component $L_z = L \cos \theta$, giving the clump 
obliquity $\theta$. The obliquity distribution was studied 
in Nesvorn\'y et al (2019). We showed that there is a $\simeq$4:1 preference for prograde rotation and that this preference is
reflected in the distribution of binary inclinations of known equal-size binaries in the Kuiper belt (Grundy et al.\ 2019). Here we focus on 
issues related to the angular momentum magnitude.

The gravity solver in {\tt ATHENA} cannot accurately model density fluctuation for wavelengths shorter that the size of the grid cells.   
To follow the gravitational collapse of clumps we lift selected clumps from {\tt ATHENA} and insert them into {\tt PKDGRAV}. The 
time of this transition, $t_{\rm A2P}$, is a model parameter. The transition cannot occur too early, when clumps have not formed yet and the 
gas effects are still important (gas is not considered in our {\tt PKDGRAV} simulations), or too late, because clumps remain 
extended in {\tt ATHENA}, interact and merge. This is unphysical because most clumps should collapse into planetesimals {\it before} 
they can interact (the collapse timescale is relatively short; Sect.\ 8.9). Ideally, we would like to approach this problem for each pebble 
clump individually by defining some physical criteria for the best transition time for that specific clump (e.g., by comparing the collision timescale 
of pebbles to their stopping time; Appendix A). For this, however, we would need to obtain complete histories of each clump with 
{\tt PLAN} and perform further analysis. This subject is left for future work. Here we adopt the same transition time $t_{\rm A2P}$ for 
all clumps in the same {\tt ATHENA} run. 

\begin{figure}
\epsscale{0.7}
\plotone{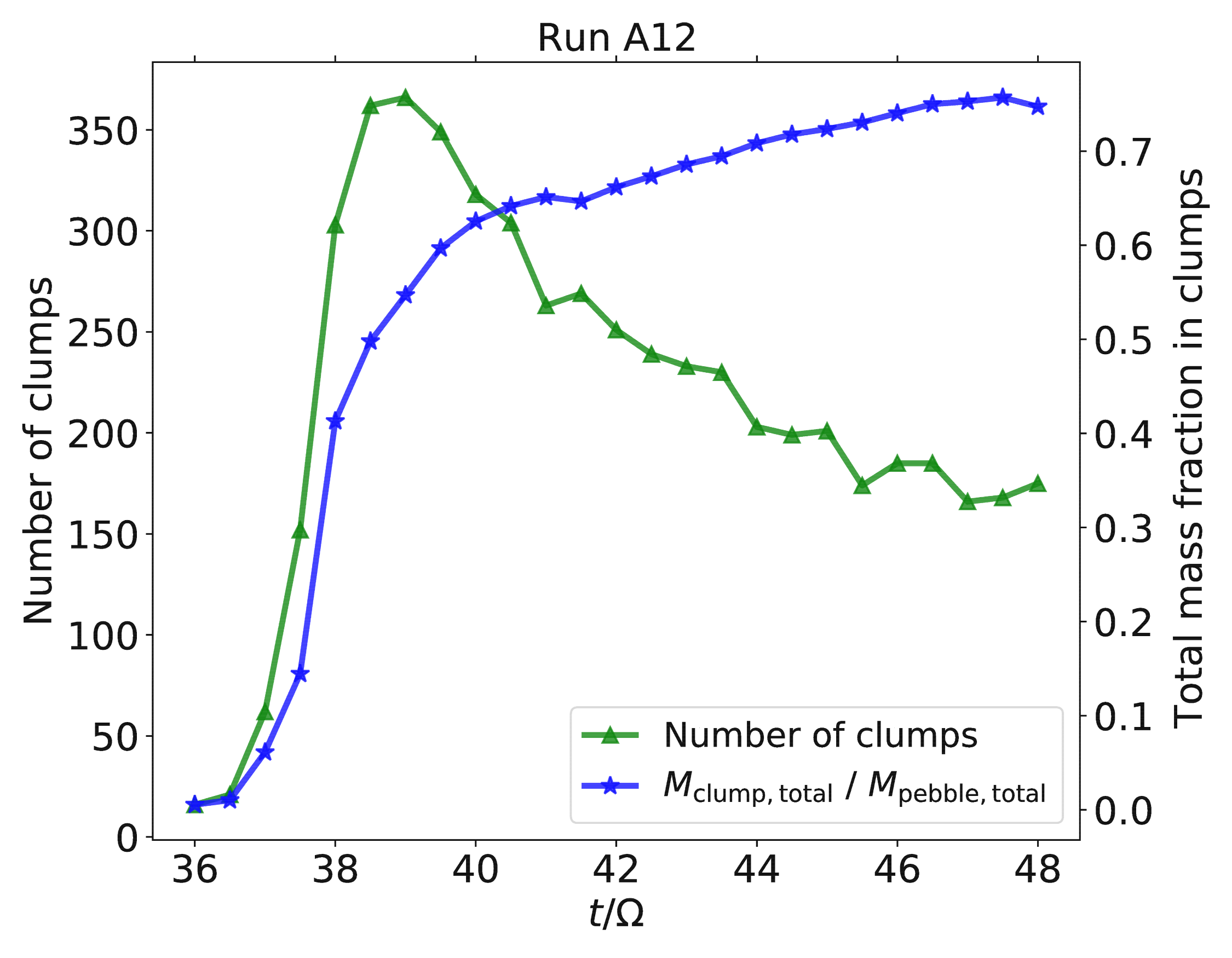}
\caption{The number of pebble clumps identified by {\tt PLAN} (green line) and the total mass fraction in clumps (blue) for the 
A12 simulation of the SI. The pebble gravity was switched on at $t=36/\Omega$. }
\label{tprof}
\end{figure}

To motivate our choice of $t_{\rm A2P}$ we plot the number of self-gravitating clumps as a function of time (Fig. \ref{tprof}). The number 
of clumps increases shortly after the pebble gravity is switched on in {\tt ATHENA} ($t_0=36/\Omega$ in A12). It reaches a maximum just past 
$t=38/\Omega$ for A12, or $t_{\rm sg}=t-t_0=2/\Omega$, and decreases over the next $\sim$$10/\Omega$. The total mass of clumps shows a two-stage 
evolution with a steep rise during the first stage ($t \Omega=36$-40 in A12) and a more gradual growth later on. 
Based on this we believe that the best choice  for $t_{\rm A2P}$ is near when the clump number reaches the maximum or near the transition 
from the first to second stages of the mass growth. We nominally set $t_{\rm A2P}=40/\Omega$ for A12 ($t_{\rm sg}=4/\Omega$) and test how the 
results change if $t_{\rm A2P}=38/\Omega$ (Sect.\ 8.5). A similar analysis was performed for B22 giving the nominal $t_{\rm A2P}=41/\Omega$ 
($t_{\rm sg}=4/\Omega$). The C203 simulation was run to $t=117.6/\Omega$ ($t_{\rm sg}=7.6/\Omega$) at which point the number of clumps 
reached a maximum. We use $t_{\rm A2P}=117.6/\Omega$ for C203. 

\section{Angular momentum}

It useful to consider the angular momentum of pebble clumps in the Hill units. This is equivalent to scaling the angular momentum of each 
clump to $l = L/L_{\rm H}$, where $L_{\rm H}=M R_{\rm H}^2 \Omega$. Here, $M$ is the clump mass, $R_{\rm H}=a (M/3M_{\rm Sun})^{1/3}$ is the 
Hill radius, $a \simeq 45$ au (Sect.\ 2), $\Omega=\sqrt{GM_{\rm Sun}}/a^{3/2}$ is the orbital frequency, $M_{\rm Sun}$ is the solar mass, and 
$G$ is the gravitational constant. The quantity $L_{\rm H}$ represents the angular momentum of a thin ring with radius 
$R_{\rm H}$ and mass $M$, rotating with the Hill velocity $v_{\rm H}=R_{\rm H} \Omega$. Intuitively, the actual angular momentum of a 
gravitationally bound pebble cloud should be smaller than $L_{\rm H}$; values $l<1$ are therefore expected.  

\begin{figure}
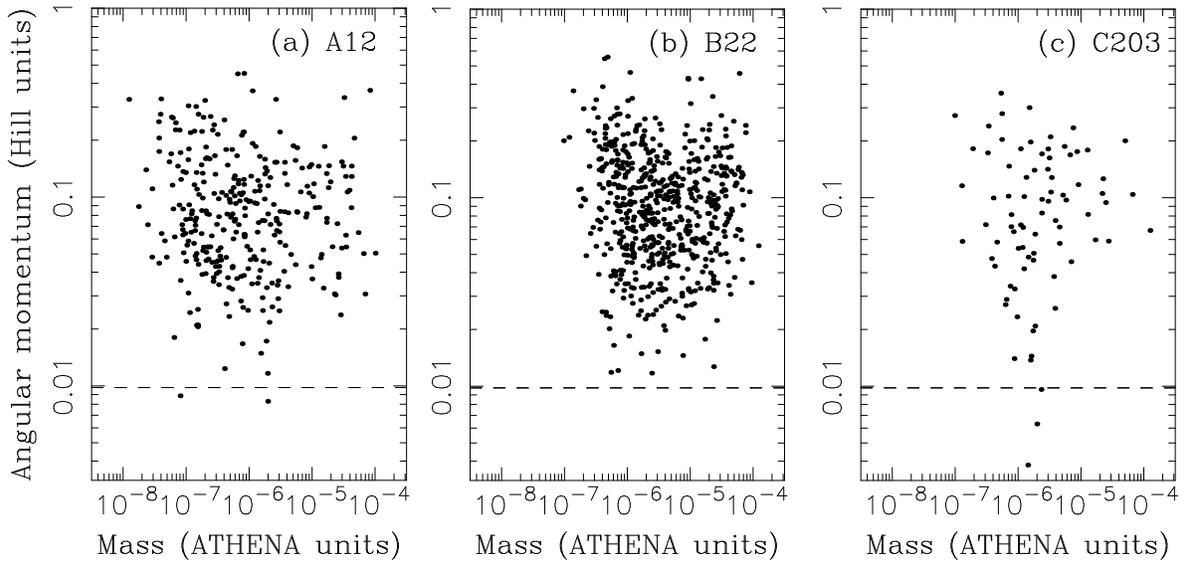

\epsscale{0.325}
\plotone{fig4a.eps}
\epsscale{0.288}
\plotone{fig4b.eps}\hspace*{3.mm}
\plotone{fig4c.eps} 
\caption{The distribution of clump mass ($M$, in the units of $M_{\rm local}$; see the main text) and SAM 
($l$) for our three SI simulations: A12 (panel a), B22 (b), and C203 (c). The clumps were extracted at $t_{\rm sg}=4 / \Omega$ in A12, 
$t_{\rm sg}=4/\Omega$ in B22 and $t_{\rm sg} = 7.6/\Omega$ in C203. The mass spectrum extends to smaller values in A12 because this simulation 
used a higher resolution (Table~2). The dashed line is the SAM value corresponding to a critically rotating Jacobi ellipsoid.}  
\label{hill}
\end{figure}

Figure \ref{hill} shows the mass scaled angular momentum (hereafter SAM) for the three SI simulations investigated here. The clump mass is given in 
the {\tt ATHENA} mass unit, $M_{\rm local} = \rho_0 H_{\rm g}^3$, where $\rho_0$ is the midplane gas density and $H_{\rm g}$ is the scale height 
of gas. Additional assumptions about the disk properties are needed to convert the results into physical mass units (Sect.~7). 
The angular momentum is computed in the inertial frame.\footnote{Eq. (8) in Nesvorn\'y et al.\ (2019) has a typo. The last term should 
have a positive sign: $+ \Omega (x_i^2 + y_i^2)$. The typo did not affect any of the results reported in that paper.}
The SAM values range between 0.01 and 0.5. The distributions extracted from different SI simulations at $t_{\rm A2P}$ are similar 
(Fig. \ref{cumul}a). In addition, the angular momentum distribution changes with 
time. It is relatively broad immediately after the first clumps form at $t_{\rm sg} \simeq 1/\Omega$ and becomes narrower later on 
(Fig. \ref{cumul}b). This happens because the outermost parts of {\tt ATHENA} clumps, which carry substantial angular momentum, 
are eroded by background gas. 

The angular momentum of clumps can be compared to that of a critically rotating Jacobi ellipsoid: $L^*=0.39 (G M^3 R)^{1/2}$ (Poincar\'e 
1885), where $M$ and $R$ are the mass and effective radius (obtained from $M$ with a reference density $\rho \simeq 1$ g cm$^{-3}$). 
When scaled to the Hill units,
$l^* = L^*/L_{\rm H} \sim 0.01$ for $a = 45$ au and the disk model discussed in Sect.\ 7. This means that typically $l/l^* \gg 1$ thus 
demonstrating that either most of the initial angular momentum must be lost or a typical SI clump cannot collapse into a large 
{\it solitary} planetesimal. The vigorous rotation of the SI clumps should be conducive to the formation of {\it binary} 
planetesimals (Nesvorn\'y et al.\ 2019).

For an exactly equal-size binary with binary semimajor axis $a_{\rm b}$ and physical radii $R_1=R_2$ ($=R$), the binary SAM
is $l_{\rm b} \simeq 0.7\, \theta^{1/2}\, (a_{\rm b}/R_{\rm b})^{1/2}$ (binary eccentricity $e_{\rm b}=0$ assumed here; a weak 
dependence on density not explicitly given), where $\theta=R_{\rm Sun}/a$ and $R_{\rm Sun} \simeq 0.0047$ au is the solar radius. So, for 
example, a typical equal-size binary in the CC population with $a_{\rm b}/R_{\rm b} \sim 50$ and $a=45$ au has $l_{\rm b} \sim 0.05$, which 
is a good fraction of the angular momentum content of the SI clumps (Figs. \ref{hill} and \ref{cumul}). This may suggest that a 
relatively large fraction of the initially available momentum is retained during collapse and used to assemble binary planetesimals.  
 
\section{{\tt PKDGRAV} simulations of gravitational collapse}

\begin{figure}
\epsscale{0.9}
\plotone{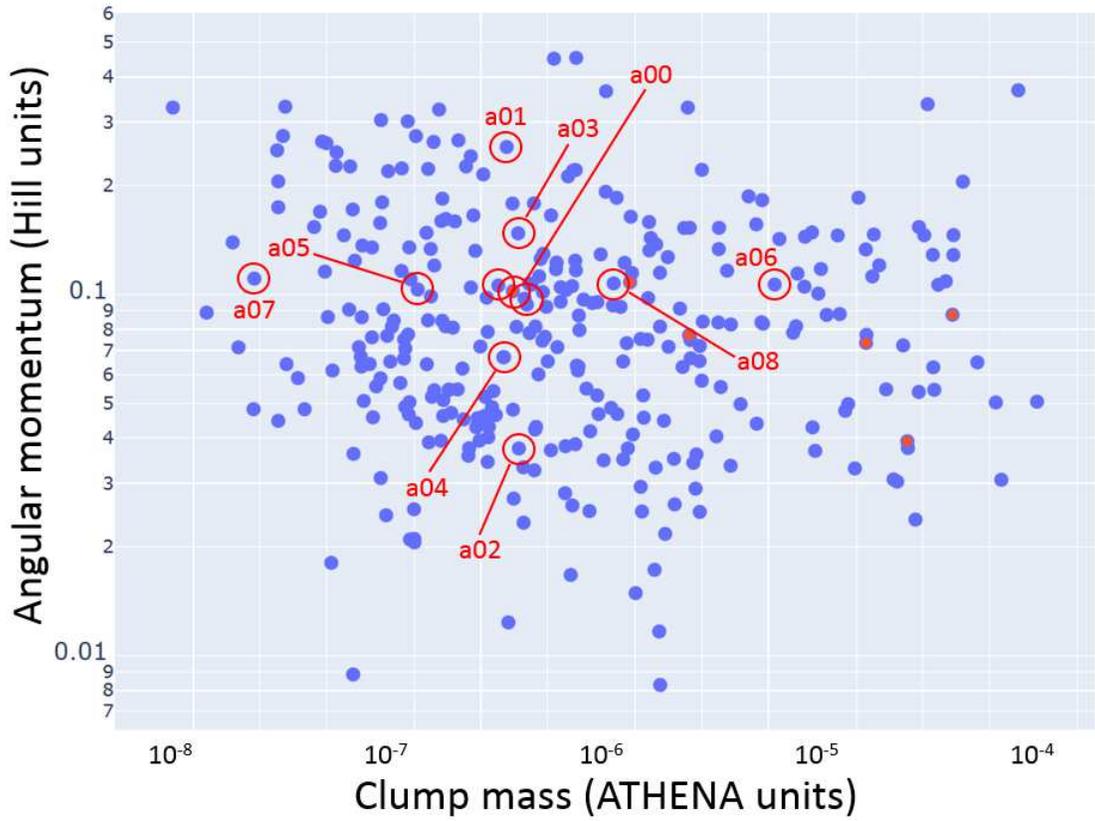} 
\caption{The SAM and mass of pebble clumps detected with {\tt PLAN} for A12 and $t=40/\Omega$. The angular momentum is
given in Hill units (Sect.\ 6) and the mass in units of $M_{\rm local}$ (Sect.\ 7). The selected clumps for the {\tt PKDGRAV} 
simulations are labeled. The a09 and a10 clumps are marked by the two circles close to a00.}
\label{marked}
\end{figure}

There are hundreds of different clumps identified in each {\tt ATHENA} simulation. Since the {\tt PKDGRAV} 
collapse simulations gobble substantial CPU resources, it is not practical, at least not in this initial study, to simulate all clumps. 
Instead, we select 8-16 clumps from each {\tt ATHENA} run using a criterion based on their mass and SAM 
(Fig. \ref{marked}). A reference clump from A12, called a00, is chosen near the middle of the mass and angular momentum 
distribution ($M/M_{\rm local}=1.4\times10^{-3}$, $l=0.102$). We then pick four additional clumps with roughly the same mass and 
larger/smaller values of the angular momentum (a01 to a04; Table 3). Additionally, four clumps are chosen with roughly the same SAM value 
as a00 but with different initial masses (a05 to a08). Finally, two clumps (a09 and a10) are selected close to the a00 clump 
to test whether clumps with the same initial mass and the same initial angular momentum produce the same planetesimal/binary 
properties. We thus have 11 clumps for A12. A similar selection is done for B22 and C203 (Table 4). We also test how the 
results are influenced by the time of transition, $t_{\rm A2P}$, from {\tt ATHENA} to {\tt PKDGRAV}. For example, all clumps
selected for run A12 at $t=40/\Omega$, except for a01, are also identified at $t=38/\Omega$ (the a01 clump 
formed after $38/\Omega$). We extract them with $t_{\rm A2P}=38/\Omega$, integrate with {\tt PKDGRAV}, and report the results in 
Sect.\ 8.5. 

\begin{figure}
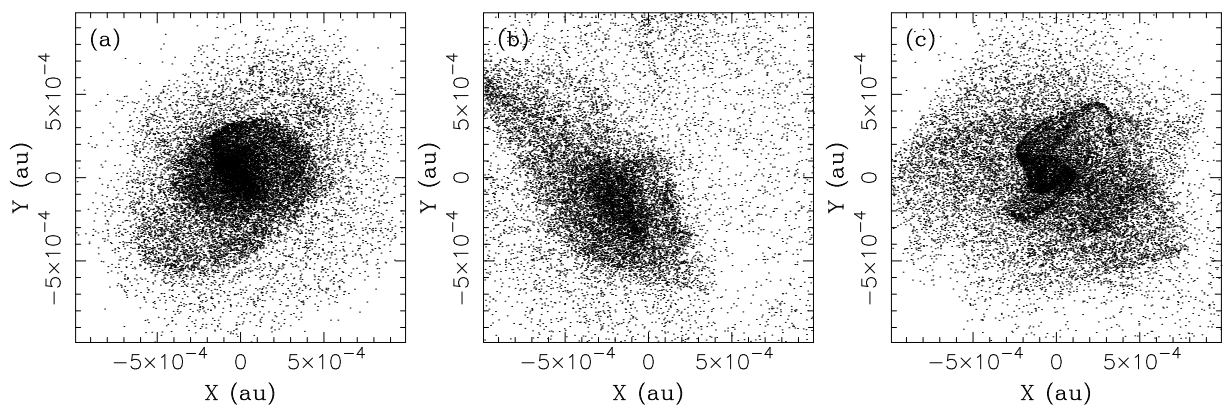

\epsscale{0.32}
\plotone{fig7a.eps}
\plotone{fig7b.eps}
\plotone{fig7c.eps} 
\caption{The initial structure of three clumps extracted from run A12: a00 (a), a01 (b) and a02 (c). The clumps are projected 
onto a plane perpendicular to their angular momentum vector. They have roughly the same mass corresponding to a sphere
with equivalent radius $R_{\rm eq} \sim 60$ km (for a reference density $\rho=1$ g cm$^{-3}$). Each box shown here is roughly 
300,000~km across. For scale, the Hill radius is $R_{\rm H} \sim 370,000$ km (for $a \simeq 45$ au).}
\label{xy}
\end{figure}

The {\tt ATHENA} simulations are done in the shearing box approximation. Since the gravitational collapse occurs on a timescale
that is much shorter than the orbital timescale, we perform the {\tt PKDGRAV} simulations in the inertial reference 
frame where the solar gravity is neglected. This greatly simplifies the analysis and visualization of the collapse results. To 
convert velocity vectors to the inertial frame, we first add the Keplerian shear which is accomplished by adding the term 
$- 1.5 \Omega x$ to the $y$ component of particle velocities (the $x$ axis points toward the Sun, the $y$ axis is in the direction 
of orbital motion; $\Omega=1$ in the {\tt ATHENA} units -- see below). Second, we remove the orbital motion around the Sun 
by applying the following transformation: $v'_x=v_x-\Omega y$, $v'_y=v_y+\Omega x$ and $v'_z=v_z$, where primed velocities are  
in the inertial frame. The origin of the reference system is then translated to the center of mass of the selected clump and 
the clump itself is rotated such that its $z$ axis points along the angular momentum vector. The initial structure of three 
selected clumps is illustrated in Fig. \ref{xy}. 

In {\tt ATHENA}, the standard gas parameters are set to unity: $\rho_0 = H_{\rm g} = \Omega = c_{\rm s} = 1$, where $\rho_0$ is the 
midplane gas density, $H_{\rm g}$ is the vertical scale height of the gas, $\Omega$ is the Keplerian frequency and $c_{\rm s}$ is the 
sound speed. This setup is useful because it circumvents the need, at least initially, for linking the gas parameters to a concrete 
disk model. To convert masses, positions and velocities into physical units ({\tt PKDGRAV} units are au, $M_{\rm Sun}$ and 
${\rm yr}/2 \pi$), however, we need to adopt some disk properties at 45 au. We use the model discussed in the 
Supplementary Information of Nesvorn\'y et al.\ (2019). In this model, the {\tt ATHENA} mass unit is  
\begin{align}
M_{\rm local} = \rho_0 H_{\rm g}^3 \simeq 2500 (T_{25} R_{45})^{3/2} \frac{\tilde{G}}{0.05}\, M_{\rm Ceres}
\end{align} 
with $M_{\rm Ceres} = 9.4 \times 10^{23}$ g. Here, the local temperature and radial distance are scaled to $T_{25} = T/(25\, {\rm K})$
and $R_{45} = r/(45\, {\rm AU})$, respectively, and $\tilde{G} = 0.05$ (Table 2). We also adopt $H_{\rm g}=0.067\, T_{25}^{1/2} r$ and 
$c_{\rm s}^2= k_{\rm B} T /(\mu m_{\rm p})$, where $k_{\rm B}$ is the Boltzmann constant, $\mu=2.34$ is the mean molecular mass, 
and $m_{\rm p}$ is the proton mass.

For each selected clump, {\tt ATHENA} particles are mapped to {\tt PKDGRAV} superparticles. 
The selected {\tt ATHENA} clumps contain $\simeq$1,500 to $\simeq$350,000 particles (Table 3). By trial and error we establish that the best 
(practical) resolution that can currently be used with {\tt PKDGRAV} is $\sim 5 \times 10^5$ SPs.\footnote{One collapse 
simulation with $5 \times 10^5$ SPs takes $\sim150$ wallclock hours to finish using pthreads on 28 Broadwell cores (all on the same node) 
of the NASA Pleiades supercomputer.} We thus increase the resolution 
in most cases. This is done by cloning each {\tt ATHENA} particle $N_{\rm cl}$ times, where $N_{\rm cl}=1$-300 depending on the number of 
{\tt ATHENA} particles in each clump (Table 3). The cloning is done by splitting every particle $N_{\rm cl}$ times and randomly 
assigning slightly different positions to the new particles (fractional change $\sim 10^{-2}$). The velocities remain unchanged. 
We verified that the cloning procedure does not 
produce artifacts. Five new distributions are generated for every clump with different random seed initializations to study 
the statistical variability of the results.

Each {\tt PKDGRAV} superparticle (SP; Sect.\ 4) is assumed to represent a large number of pebbles with nominal radii $r_{\rm peb}=1$ 
cm (we also tested other pebble sizes and report the results in Sect 8.3).  
Colliding SPs inelastically bounce, with fixed $C_{\rm R}$ value, and are merged if the 
collision speed becomes lower that the mutual escape speed. If that's the case, two SPs are combined into a new SP following the method 
described in Sect.\ 4. Particle merging greatly increases the speed of {\tt PKDGRAV} simulations and reduces problems with packing
of inflated SPs (as described in Nesvorn\'y et al.\ 2020).\footnote{This means that our {\tt PKDGRAV} runs do not yield any shape information
because SPs remain spherical. Studies of planetesimal shape, which would be relevant for Arrokoth (Stern et al.\ 2019, McKinnon et al. 
2020), are left for future work.} Each cloud is considered in isolation (i.e., there is no interaction between clumps in our {\tt 
PKDGRAV} runs) and the gas effects are neglected (Appendix A). The collapse simulations are run over 100 yr, roughly 1/3 of the orbital 
period at 45 au, at which point planetesimals and planetesimal binaries have already formed (Sect.\ 8.8). We checked that
the binary properties do not change much (fractional change $\sim 1$\%) from $t=50$ to 100 yr thus demonstrating that the integration time is 
generally adequate (Robinson et al. 2020). In several cases, where hierarchical systems with multiple moons formed, we continued 
the simulations to $t=1000$ yr. These cases are discussed in Sects. 8.4 and 8.9. If, as it often happens, the SPs are still inflated at 
the end of the simulations, we set $n=1$ (Sect. 4) and recalculate the planetesimal radii for the reference density $\rho=1$ g cm$^{-3}$. 

\section{Results} 

Here we discuss the results of our collapse simulations. We combine the {\tt ATHENA} and {\tt PKDGRAV} results to determine the properties 
of planetesimals and binaries that are expected to form for adopted disk conditions, and characterize the dependence of the results on model 
parameters such as $C_{\rm R}$, $f$, pebble size, etc. To identify bound binaries in the simulations we first sort bodies by size. 
The code then goes down the list, from the largest to smallest bodies, and checks whether the binding energy of each pair is negative 
and, if so, whether the semimajor axis of the binary orbit is smaller than the Hill radius. It then proceeds, again from large to 
small, to consider triple and multiple systems, and checks on the binding energy of each combination. The details of this method are 
similar to that used in Robinson et al. (2020).

\subsection{$C_{\rm R}$ and $f$}

We simulate gravitational collapse of pebble clump a00, extracted at $t_{\rm A2P}=40/\Omega$, to understand the dependence of the results 
on the coefficient of restitution $0 < C_{\rm R} \leq 1$ and the sampling reduction factor $1 \leq f < 2$. Specifically, we set $C_{\rm R}=0.1$, 
0.3, 0.5, 0.7, 0.9 (for context, note that NYR10 used $C_{\rm R}=0$) and $f=1$, 1.25, 1.5 and 1.75 in different runs. We find that, if 
$C_{\rm R} = 0.9$, the particle collisions are nearly 
elastic and this leads to a situation where the accretion growth within the collapsing cloud is relatively modest. Eventually, when a 
single massive body forms in the center, it is surrounded by a 
large envelope of smaller objects where the angular momentum is stored. The envelope should presumably become flatter over time, as energy 
is slowly removed by collisions, but we would need to extend the simulations to see that actually happening. 

\begin{figure}
\epsscale{0.7}
\plotone{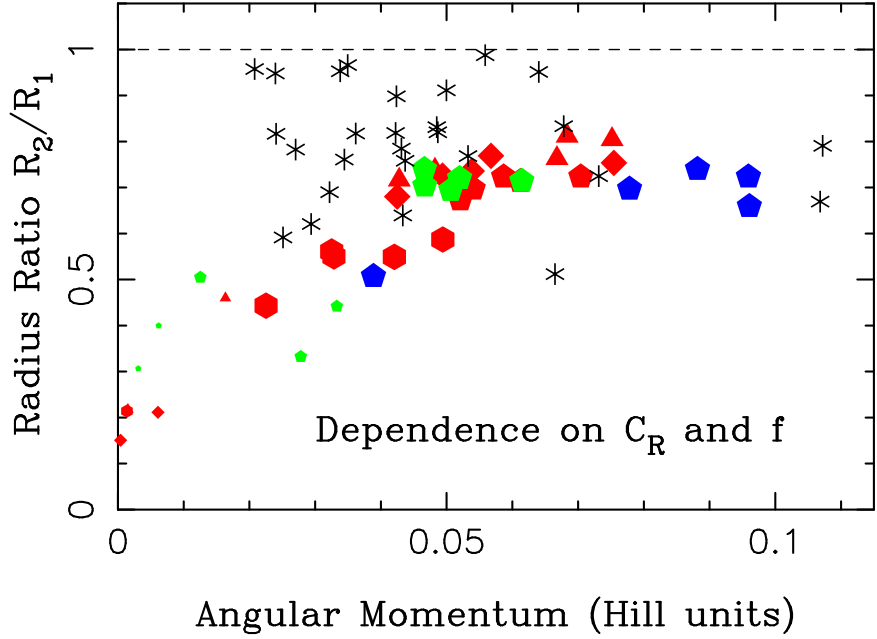}
\caption{The physical properties of planetesimal binaries and multiple systems that formed in the a00 clump collapse. 
Pebbles were assumed to have the initial radius $r_{\rm peb}=1$ cm. The shape of 
a symbol indicates the value of the coefficient of restitution used in the {\tt PKDGRAV} simulations: $C_{\rm R}=0.1$ (triangle),  
$C_{\rm R}=0.3$ (square), $C_{\rm R}=0.5$ (pentagon), $C_{\rm R}=0.7$ (hexagon). The results for $C_{\rm R}=0.9$ are not shown
here because the simulations with near-elastic collisions do not produce planetesimal binaries. Color indicates different 
sampling reduction factors: $f=1.5$ (red), $f=1.75$ (green), and $f=1$ (blue). The largest binary components are plotted  
with the largest symbols; the symbol size is reduced for moons in higher multiplicity systems. The smallest satellites 
considered here have $R\sim10$ km. For reference, the CC binaries from Table 1 are plotted with asterisks. Two known binaries, 
2000 CF105 with $l_{\rm b} \simeq 0.15$ and 2001 QW322 with $l_{\rm b} \simeq 0.18$, fall outside the plotted range.}  
\label{restit}
\end{figure}

The equal-size binaries with $R_2/R_1 > 0.5$ form for $C_{\rm R} \leq 0.7$ (Fig. \ref{restit}). In this case, the mass of the final 
binary represents $>90$\% of the initial a00 mass meaning that the mass loss is relatively small (the mass loss correlates 
with the initial angular momentum; Table 3). For $C_{\rm R} \leq 0.5$, the model binaries are more equal sized ($R_2/R_1 \simeq 0.7$) 
than for $C_{\rm R} = 0.7$ ($R_2/R_1 \simeq 0.5$). Interestingly, the collapse of a00 does not produce binaries with $R_2/R_1 > 0.85$ 
for any choice of $C_{\rm R}$ (and $f$), whereas these strictly equal-size cases represent a large share of known CC binaries 
(8 out of 29 shown in Fig. \ref{restit}; 28\%). We find below that more equal-sized binaries form in simulations with an 
improved resolution (i.e., higher number of SPs) and/or with larger pebbles (here $r_{\rm peb}=1$ cm).

The SAM of model binaries is $0.02<l<0.1$, reflecting a range of binary separations. The final value of SAM correlates both with 
$C_{\rm R}$ and $f$. The results for $0.1 \leq C_{\rm R} \leq 0.5$ are similar ($0.04<l<0.08$), but
those with $C_{\rm R} = 0.7$ indicate a larger angular momentum loss (final $0.02<l<0.05$; a00 had $l_0=0.102$ initially). The 
case with $f=1$ leads to more efficient accretion with a larger fraction of the initial angular momentum being stored in 
the final binary (75-95\%; blue symbols in Fig. \ref{restit}), except for one case where an unequal-size binary formed with 
$l \simeq 0.04$. There is not much difference between the results obtained with $f=1.5$ and $f=1.75$; both these choices lead to 
binaries with $R_2/R_1 \simeq 0.5$-0.7 and $l \simeq 0.03$-0.07. The latter corresponds to $\sim$30-70\% of the initial SAM value, which is 
significant because this could explain why the known CC binaries (Sect.\ 8.7) have, on average, lower SAM values than the SI clouds
(i.e., angular momentum is lost during collapse). We use $C_{\rm R}=0.5$ and $f=1.5$ in the following investigation.

\subsection{Resolution}

Figure \ref{result}a shows the dependence of the results on resolution. To make this figure, we simulated clump a00 
with $C_{\rm R}=0.5$, $f=1.5$ and  $r_{\rm peb}=1$ cm, and varied the number of SPs. Clump a00 has 22788 {\tt ATHENA} particles 
at $t=40/\Omega$. A different number of clones was used to set up different {\tt PKDGRAV} simulations: $N_{\rm cl}=1$, 3, 10 and 
30. The simulation with the best resolution thus have nearly 700,000 SPs. [We were unable to complete runs with even higher 
resolution because the {\tt PKDGRAV} code dramatically slows down as the number of SPs approaches $10^6$. This happens due
to a large number of collisions during the disk stage (Sect.\ 8.8).] 

When the number of SPs is boosted by a factor of $f_{\rm n}$, the sampling rate drops by the same factor (i.e., each SP initially 
represents fewer pebbles). Since the SP radius is $R \propto \sqrt{n} r_{\rm peb}$, where $n$ is the number of pebbles it represents 
(Sect.\ 4), the SP size is proportional to $f_{\rm n}^{-1/2}$. So, for example, the SPs are $\sqrt{10}\simeq3.2$ smaller in the case 
with $N_{\rm cl}=10$ than for $N_{\rm cl}=1$. 

The runs with higher resolution produce binary planetesimals with more equal sized components (i.e., higher $R_2/R_1$) and 
these binaries tend to retain more of the initial angular momentum (Fig. \ref{result}a). For example, we find $R_2/R_1=0.4$-0.5 
for $N_{\rm cl}=1$ and $R_2/R_1=0.7$-0.8 for $N_{\rm cl}=10$. The highest-resolution simulations with $N_{\rm cl}=30$ show 
a slightly larger spread of $R_2/R_1$, with the maximum value $R_2/R_1=0.96$, and $l=0.07$-0.09. The results with $N_{\rm cl}=3$ 
are intermediate. This suggests that the interaction of SPs in the low resolution cases can lead to unbalanced accretion where 
one of the binary components grows larger than it should. The resolution dependence may explain the problem pointed out in the 
previous section, where none of the intermediate-resolution cases produced binaries with $R_2/R_1 > 0.85$ (whereas the CC 
binaries with $R_2/R_1 > 0.85$ are common). 

\subsection{Pebble size}

The a00 clump collapse was simulated with $r_{\rm peb}=1$ mm, 1 cm and 10 cm to understand the dependence of the results on 
pebble size (Fig. \ref{result}b). To isolate the effects of pebble size from those of the resolution, these simulations used the 
same cloning factor, $N_{\rm cl}=10$, and therefore the same number of SPs. As the pebble size increases (decreases) by a factor 
of $f_{\rm r}$, but the number of SPs remains the same, each SP must represent a smaller (larger) number of pebbles. Since
the SP radius is $R \propto \sqrt{n} r_{\rm peb}$ (Sect.\ 4), the SP size is proportional to $f_{\rm r}^{-1/3}$. This means that 
the initial inflation of SPs in the {\tt PKDGRAV} code modestly decreases (increases) for $f_{\rm r}>1$ ($f_{\rm r}<1$). The effects 
of larger pebble size are therefore in some sense similar to those of improved resolution (except that increasing the 
pebble size also leads to fewer collisions and lower energy loss).

We find that the a00 clump simulations with larger pebbles lead to binaries with components that are more similar in size (Fig. 
\ref{result}b) -- this trend was confirmed for other pebble clumps as well. 
Specifically, for a00, $R_2/R_1=0.5$-0.6 for $r=1$ mm and $R_2/R_1=0.8$-1 for $r=10$ cm. 
The dependence of the results on pebble size and resolution is an expression of the same underlying trend. 
In both cases, more equal-size binaries form when SPs are smaller and allow for more balanced accretion on 
binary components. 
This highlights the importance of convergence studies. Here we content ourselves with pointing out that the identified trend 
goes in the right direction in that the simulations with smaller SPs are expected to yield $R_2/R_1 > 0.85$ more often. 

\subsection{Different clumps}

\begin{figure}
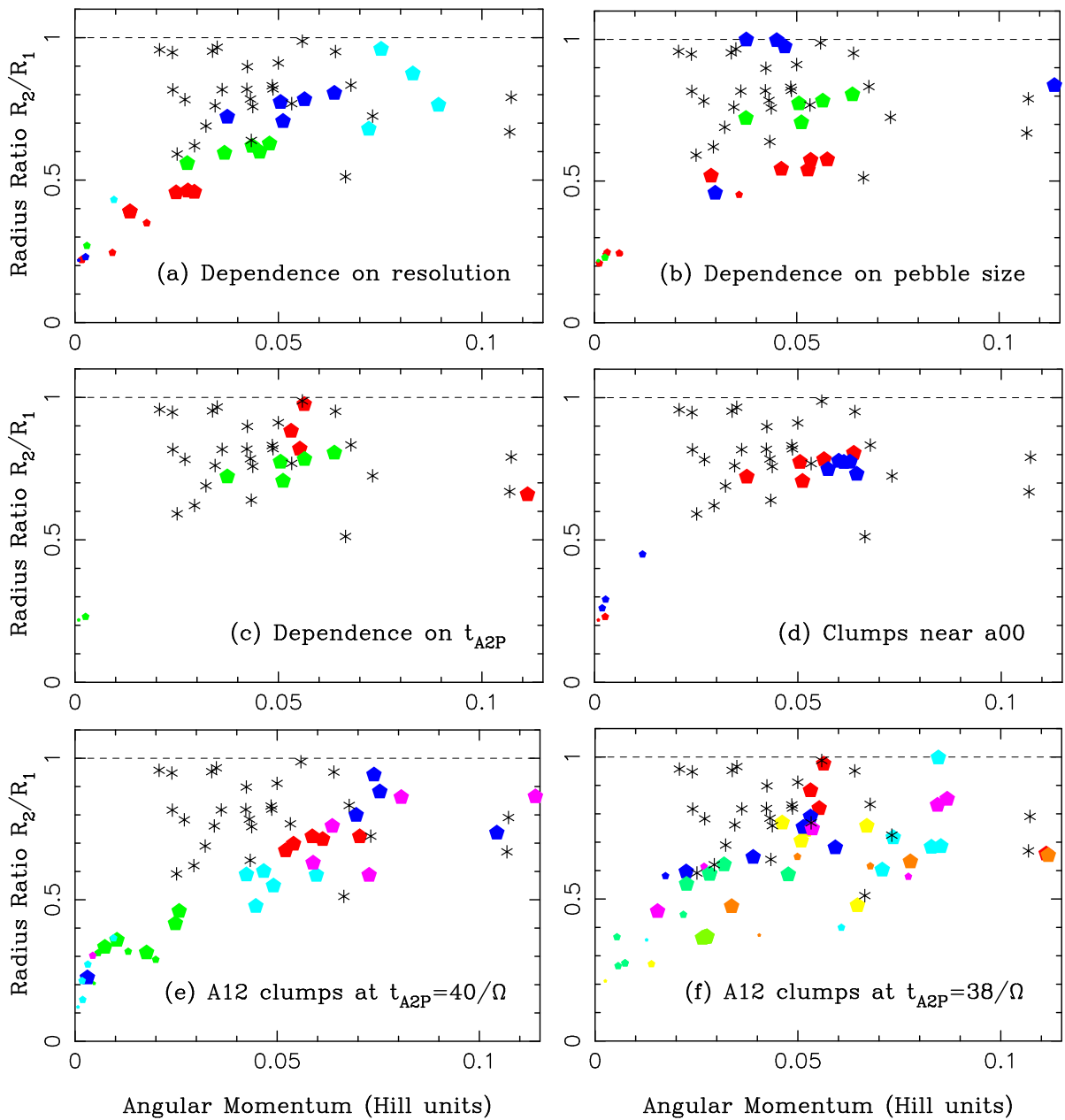

\epsscale{0.49}
\plotone{fig9a.eps}
\epsscale{0.456}
\plotone{fig9b.eps}\\[2.mm]
\epsscale{0.49}
\plotone{fig9c.eps}
\epsscale{0.456}
\plotone{fig9d.eps}\\[2.mm]
\epsscale{0.49}
\plotone{fig9e.eps}
\epsscale{0.455}
\plotone{fig9f.eps}
\caption{\small The physical properties of binaries and multiple systems obtained in the {\tt PKDGRAV} simulations with 
$C_{\rm R}=0.5$ and $f=1.5$. The results for the a00 pebble clump are shown in panels a to c.
Panel a -- Color indicates cases with different cloning factors: $N_{\rm cl}=1$ (red), $N_{\rm cl}=3$ (green), 
$N_{\rm cl}=10$ (blue), and $N_{\rm cl}=30$ (turquoise). 
Panel b -- Color indicates different pebble sizes: $r_{\rm peb}=1$ mm (red), $r_{\rm peb}=1$ cm (green), and 
$r_{\rm peb}=10$ cm (blue). 
Panel c -- Color indicates different transition times from {\tt ATHENA} to {\tt PKDGRAV}: $t_{\rm A2P}=38/\Omega$ (red) 
and $t_{\rm A2P}=40/\Omega$ (green). 
Panel d -- different pebble clumps: a00 (red) and a10 (blue). Clump a09 with the initial parameters near 
a00/a10 did not produce binary planetesimals. 
Panel e -- clumps selected with $t_{\rm A2P}=40/\Omega$: a00 (red), a01 (green), a03 (blue), a06 (turquoise) and a08 (purple). 
Panel f -- clumps selected with $t_{\rm A2P}=38/\Omega$: a00 (red), a02 (yellow), a03 (blue), a05 (orange), a06 (turquoise), 
a07 (light green), a08 (purple) and a10 (dark green). See the caption of Fig. \ref{restit} for additional information.}  
\label{result}
\end{figure}

Figure \ref{result}e shows the results of collapse simulations for pebble clumps extracted from run A12 with $t_{\rm A2P}=40/\Omega$. 
Overall, the binaries produced in these simulations are similar to the observed CC binaries, at least in terms of 
the radius ratio ($R_2/R_1=0.5$-1) and SAM ($l=0.04$-0.12). In addition to the equal size binaries 
with $R_2/R_1>0.5$ some clumps produce unequal-size binaries and multiple systems with $R_j/R_1=0.1$-0.5, where 
index $j \geq 2$ denotes secondaries and smaller moons. A good example of this is the a01 clump that started with 
roughly the same mass as a00 but with large SAM. There are up to four moons with $R_j>10$~km and many smaller ones (green 
symbols in Fig. \ref{result}e). We tracked these systems longer to test their stability and found that some of them are
stable whereas others are reduced to one or two moons by $t=1000$ yr. The long-term stability of multiple systems 
has yet to be investigated.

Clump a03 with roughly the same mass as a00 and a01, and intermediate angular momentum, produces two binaries with
$R_2/R_1>0.85$ (blue symbols in Fig. \ref{result}e; out of five runs with different seeds, $\sim$40\%). These strictly 
equal-size binaries were difficult to obtain in the simulations of a00. For a00, the strictly equal-size binaries  
were obtained only when the resolution or pebble size were increased (Sects. 8.2 and 8.3). We therefore see that 
there are multiple pathways to resolving the issue with the strictly equal-size binaries.
     
Some clumps, namely a02, a04, a05 and a07, do not produce any gravitationally bound systems of bodies with $R_2>10$ km. 
The simulations of these clumps end up with a large planetesimal in the center of the collapsed cloud and an
envelope of $R<2$ km bodies extending to the edge of the Hill sphere.  The clumps in question fall into two categories. 
Clumps a02 and a04 have roughly the same mass as a00 but low initial SAM. These clumps are not expected to produce 
well-separated binaries with large angular momentum. They may produce close or contact binaries but our integration method 
with the inflated SPs cannot resolve them, or they may not produce binaries at all. This issue can be addressed 
by increasing the resolution (see below), using higher sampling reduction factors and/or larger 
pebbles.\footnote{The degree to which these results are affected by the limitations of the gravity solver in {\tt ATHENA} is not 
clear. There are at least two issues. The first one is that the clumps stay extended and can interact in {\tt ATHENA} whereas this interaction 
should be minimal in reality because clumps collapse into planetesimals on a short timescale (Sect.\ 8.8). The second one is
that the internal structure of each clump is affected by the {\tt ATHENA} gravity solver. This problem is more important for 
smaller clumps because their Hill sphere is more poorly resolved by {\tt ATHENA}'s grid cell. The cell sizes are 0.0002 in A12, 
and 0.0004 in B22 and C203 (all in the units of $H_{\rm g}$). For comparison, the Hill radius of a00 is 0.00083 such that there 
are roughly eight cells over the whole Hill sphere. The least-massive clump investigated here, a07, has $R_{\rm H} \simeq 
0.00034$ and there are only $\sim$3.5 cells covering the whole Hill sphere, which may not be adequate. We find that small and 
large clumps behave differently in the {\tt ATHENA} simulations. The small clumps remain more extended, often over their whole Hill 
sphere, probably because the gravity solver in {\tt ATHENA} does not allow them to contract. The massive clumps evolve 
to become more compact in {\tt ATHENA}. This may influence our {\tt PKDGRAV} results. A more accurate approach to this problem, 
where different {\tt ATHENA} clumps are continuously tracked over time and $t_{\rm A2P}$ is set for each of them individually, 
is left for future work.}

\begin{figure}
\epsscale{0.7}
\plotone{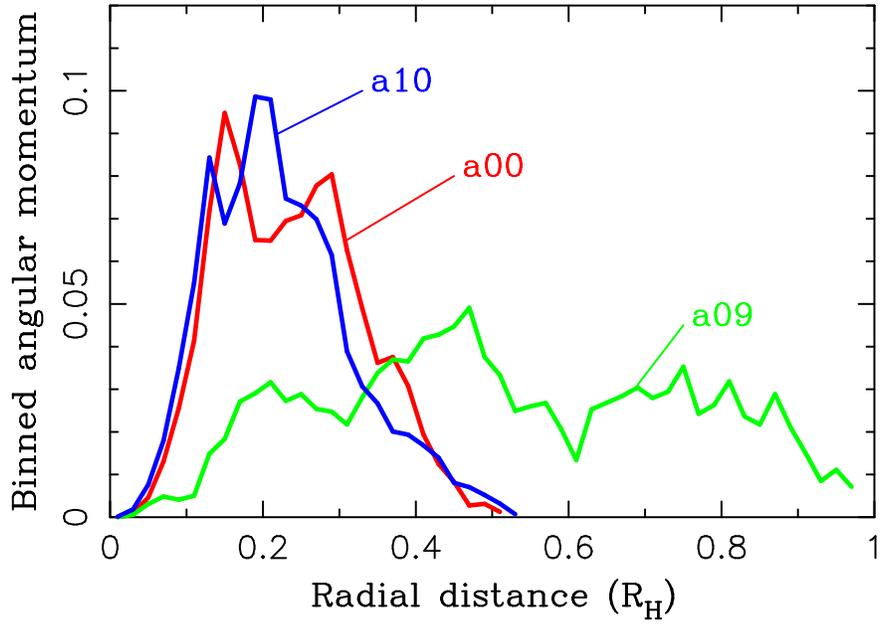}
\caption{The initial radial distribution of the angular momentum for three different clumps (A12 and $t_{\rm A2P}=40/\Omega$):
a00 (red line), a09 (green) and a10 (blue). }  
\label{profiles}
\end{figure}

Clumps a05 and a07 have roughly the same SAM as a00 but lower mass. If the SAM were the only parameter that determines 
whether binaries form, equal-size or not, it would seems strange that a05 and a07 do not produce any. {\it Here we find that 
binary formation depends on the radial SAM distribution}. To demonstrate this, we simulate two additional 
clumps (a09 and a10; Table 3) with initial parameters similar to the a00 clump, and find that a10 gives results similar to 
a00 (Fig. \ref{result}d) but a09 does not produce any binaries. This turns out to be a consequence of the 
initial radial profile (Fig. \ref{profiles}). In the two clumps that produce equal-size binaries, a00 and a10, 
nearly all initial SAM is contained in a sphere with radius $\simeq 0.3 R_{\rm H}$. In contrast, clump a09 shows a more 
extended distribution. We return to this issue in more detail in Sect.\ 8.6.

The CC binaries shown in Fig. \ref{result}e populate the region with $R_2/R_1>0.8$ and $l<0.05$ but the model 
binaries do not. This issue may be related to the resolution of our collapse simulations, because the equal-size binaries 
with low angular momentum have tight orbits that are the most difficult to resolve with our SP method 
(binary components may artificially be merged). Alternatively, it can be a consequence of our hand-over method from {\tt ATHENA} 
to {\tt PKDGRAV}. As we explained in Sect.\ 5, we adopted a crude approximation with $t_{\tt A2P}=40/\Omega$ for all clumps. 
If we choose an earlier transition time,  $t_{\tt A2P}=38/\Omega$, the angular momentum content in model binaries
is slightly reduced (Sect.\ 8.5; Fig. \ref{result}f). This may suggest that an even earlier transition, done individually 
just after each clump forms, could produce a better match to observations. 
 
\subsection{Transition time $t_{\rm A2P}$}

So far we discussed the results obtained with $t_{\rm A2P}=40/\Omega$. Here we consider an earlier transition time,
$t_{\rm A2P}=38/\Omega$, in detail. We do not select a new set of pebble clumps at $t=38/\Omega$. Instead, we track the previously 
selected clumps --a00 to a10-- from $t=40/\Omega$ to $38/\Omega$. This is done by following individual particles, 
which have universal identifiers in {\tt ATHENA}, from $t=40/\Omega$ to $38/\Omega$ and checking whether these 
particles belong to any clump identified by {\tt PLAN} at $t=38/\Omega$.  We find that all clumps selected at 
$t=40/\Omega$, except for a01, already exist at $t=38/\Omega$ (Table~3). The a01 clump has the largest SAM value, 
$l=0.256$, of all selected clumps at $t=40/\Omega$. This, in itself, is an indication that it has formed just shortly 
before $t=40/\Omega$, because there is a tendency for clumps being born with a relatively high angular momentum 
(Fig. \ref{cumul}b).

The SAM value of any individual clump changes substantially from $t=38/\Omega$ to $40/\Omega$ (Table~3). In some cases, such 
as a04 and a06 with the highest and lowest SAM values at $t=38/\Omega$, respectively, we find that {\tt PLAN} annexed
two nearby clouds under the same identification. The two clouds merged after $t=38/\Omega$ and became the more massive 
a06 clump identified at $t=40/\Omega$. In all other cases, the selected clumps are isolated at $t=38/\Omega$. 
Yet, when we compute the individual clump's SAM at different times we see that it changes. It remains to 
be understood whether this happens because of mutual clump interaction, aerodynamic gas effects, or some 
other process. The masses of individual clumps at $t=38/\Omega$ and $40/\Omega$ are often similar (a00, a03, a05, a08, a09 
and a10 show $<20$\% change in the total mass). An extreme mass change happens for a07 which has $M \simeq 1.3 \times 
10^{-6}$ $M_{\rm local}$ at $t=38/\Omega$ and $M \simeq 2.9 \times 10^{-8}$ $M_{\rm local}$ at $t=40/\Omega$. It thus seems 
that a07, as identified at $t=40/\Omega$, is just a small segment of a much larger structure found at $t=38/\Omega$. 
Indeed, by inspecting different time frames from the {\tt ATHENA} simulation, we found that a07 at $t=40/\Omega$ is a 
tidally disrupted fragment of the a07 clump identified at $t=38/\Omega$.

We first simulate collapse of a00, extracted with $t_{\rm A2P}=38/\Omega$, to see whether the time of transition 
makes any difference. We find that the results are broadly similar but the earlier transition resulted in two
strictly equal-mass binaries with $R_2/R_1>0.85$, one binary with a relatively large value of SAM, $l=0.11$,
and one case where no binary formed (Fig.~\ref{result}c). It thus seems that the earlier transition time for a00 
leads a broader spectrum of results (probably related to the large SAM value of a00 at $t=38/\Omega$). This 
also shows that the clumps with a large SAM content do not necessarily lead to hierarchical systems of planetesimals; 
they can form equal-size binaries as well if the radial SAM distribution is sufficiently compact (Sect.\ 8.6). Figure 
\ref{result}f shows the results of our {\tt PKDGRAV} simulations for all clumps and $t_{\rm A2P}=38/\Omega$. 

To this point we have focused on equal-size binaries with $l=0.02$-0.08. But there are also four known CC binaries with 
$l>0.1$. Two of them have extremely large separations, 2000 CF105 with $a_{\rm b}/R_{\rm b} \simeq 600$ and 
$l\simeq 0.15$, and 2001 QW322 with $a_{\rm b}/R_{\rm b} \simeq 800$ and $l\simeq 0.18$. They fall outside the range 
of Fig. \ref{result}. None of the collapse simulations performed here produced such extremely wide binaries (our 
record value is $l_{\rm b} \simeq 0.13$; Fig. \ref{result}b). The other two binaries, 2003 UN284 and 2006 BR284, have 
$l_{\rm b} \simeq 0.107$. We find it interesting that at least some of our simulations produce these widely separated 
binaries (panels b, c, e and f in Fig. \ref{result}b). By investigating this issue in more detail, we find that 
binaries with $l_{\rm b}>0.1$ are typically a by-product of exchange reactions among three or more large planetesimals 
that form in the same pebble cloud (Section 8.8).

\subsection{Binary formation criterion}

The total SAM value is not the only parameter that matters when it comes to binary formation. This is most easily 
illustrated by comparing a00/a10 with a09. These clumps have roughly the same total SAM value (and mass) at 
$t=40/\Omega$, but whereas a00/a09 produce the equal-size binaries (Fig. \ref{result}d), clump a09 does not (at least 
with the nominal resolution; see below). This is clearly related to the radial distribution of SAM  
(Fig. \ref{profiles}). The clumps with central concentrations of SAM, such as a00 and a10, produce wide/equal-size 
binaries. The ones where the SAM profile is flat, such as a09, do not. This trend can be used to formulate binary 
formation criteria that could tell us whether a wide/equal-size binary is expected to form from a given clump, or not.

Figure \ref{criter} illustrates one such criterion that works pretty well. To make this plot we computed the SAM
content in a sphere of radius $r=0.5$ $R_{\rm H}$, where $r$ is measured from the clump center of mass. This quantity
is denoted by $l_{0.5}$ in the following text. We find that the wide/equal-mass binaries form if $l_{0.5}>0.07$ 
and do not form if $l_{0.5}<0.07$. For  $t_{\rm A2P}=38/\Omega$, the only clear exception to this rule is the a06 clump 
(blue star in  Fig. \ref{criter}), which is a composite of two pebble concentrations at $t=38/\Omega$. The criterion 
is not expected to apply to such cases. There is also a01 with $l_{0.5}=0.096$ for $t_{\rm A2P}=40/\Omega$, which leads 
to a hierarchical system with two or three large satellites with $R_j=0.19$-0.36. The a01 clump has the largest SAM
value of all clumps selected at $t_{\rm A2P}=40/\Omega$.   

It remains to be understood whether the criterion outlined above aspires to a universal rule for the formation of 
planetesimal binaries, or whether it is merely a reflection of our SP method that may be unable, at least in some 
cases, to resolve tight binaries. Assuming the former, we could roughly estimate the binary fraction expected from A12. 
For that we simply count all clumps with $l_{0.5}>0.07$ and find that this condition is satisfied in 47\% of cases 
for $t_{\rm A2P}=38/\Omega$ and 39\% of cases for $t_{\rm A2P}=40/\Omega$. This means that the wide/equal-size binaries 
should represent 39-47\% of the total population of planetesimals (if run A12 is taken as a guide). For comparison, 
Noll et al.\ (2008) estimated the binary fraction to be $>$30\% in the CC population, whereas Fraser et al.\ (2017) 
suggested it can be as high as 100\% (cf. Noll et al. 2020).

The resolution issue can be tested by increasing the number of SPs in {\tt PKDGRAV}. We reported the results of some 
of these tests in Sect.\ 8.2 and Fig. \ref{result}a. Here we perform an additional test for clump a09 that has 
$l_{0.5}=0.061$ at $t=40/\Omega$ and is therefore not expected to produce wide/equal-size binaries. For this test,
we used $t_{\rm A2P}=40/\Omega$ and increased the cloning factor to $N_{\rm cl}=30$ (from the original $N_{\rm cl}=10$).
This gives nearly 600,000 SPs, which is at the limit of our present computational resources. A tight/equal-size 
binary forms in this high-resolution simulation with $R_2/R_1=0.729$ and $l_{\rm b}=0.0326$. This binary would fall 
right in the middle of the binary-parameter domain in Fig. \ref{result}e that was left empty in our lower-resolution 
simulations. We thus see that, indeed, the {\tt PKDGRAV} resolution is limiting our capability to reproduce the 
tight/equal-size binaries. The binary formation criterion developed here should be considered with this caveat 
in mind.

\begin{figure}
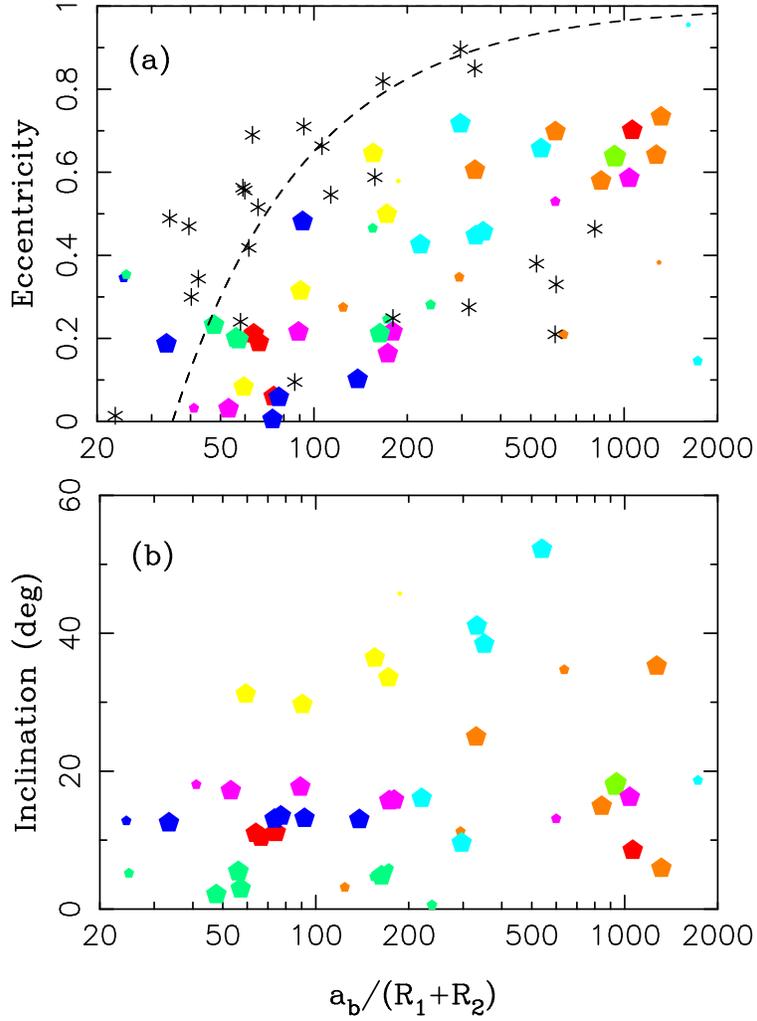

\epsscale{0.6}
\plotone{fig12a.eps}\\[2.mm]
\plotone{fig12b.eps}
\caption{Binary eccentricities (panel a) and inclinations (panel b) for A12 and $t_{\rm A2P}=38/\Omega$: a00 (red), 
a02 (yellow), a03 (blue), a05 (orange), a06 (turquoise), a07 (light green), a08 (purple) and a10 (dark green). The 
largest binary components are plotted  with the largest symbols; the symbol size is reduced for moons in higher 
multiplicity systems. For reference, the CC binaries from Table 1 are plotted with asterisks in panel (a). The dashed 
line in panel (a) denotes the pericenter distance $q_{\rm b}=a_{\rm b} (1-e_{\rm b})=35(R_1+R_2)$. The results for 
A12 and $t_{\rm A2P}=40/\Omega$, B22 and C203, not shown here, are similar. We show the results for A12 and 
$t_{\rm A2P}=38/\Omega$ because a broader number of binary outcomes are available in this case.}    
\label{ei38}
\end{figure}

Figure \ref{ei38} shows the eccentricities and inclinations of binary orbits obtained for A12. The binary inclination
is measured with respect to the initial angular momentum vector of each cloud (or, equivalently, relative to the clump 
obliquity). We found that the inclinations are generally small with over 70\% of cases showing $i<20^\circ$. This means that the 
initial obliquity distribution of clumps is a good proxy for the expected inclination distribution of binaries that
form from the clumps. We previously used this argument to show that the SI can explain the inclination distribution
of known equal-size binaries (Nesvorn\'y et al.\ 2019). The model distribution of eccentricities is broad and similar
to that of observed binaries, except that we do not obtain binaries with pericenter distance $q_{\rm b}=a_{\rm b}
(1-e_{\rm b})<35(R_1+R_2)$, whereas these orbits are common in the observed sample. This problem is a consequence of 
the insufficient resolution of our SP simulations (tight binaries are not resolved). 

\subsection{Angular momentum distribution} 

Here we compare the angular momentum distribution of model and observed binaries in more detail. The SAM distribution
for model binaries is computed by the following method. We consider the results of the simulations obtained for A12, 
$t_{\rm A2P}=38/\Omega$, $C_{\rm R}=0.5$, $f=1.5$ and $r_{\rm peb}=1$ cm. To do this rigorously we would need to simulate
every A12 clump that is detected at $t_{\rm A2P}=38/\Omega$ and determine the SAM distribution of all binaries. But we 
do not have that information available. To construct an approximate SAM distribution, we consider one clump at 
a time and find its nearest neighbor in the $(l,l_{0.5})$ plane that we simulated with {\tt PKDGRAV} (stars in Fig. \ref{criter}a). 
We then use the {\tt PKDGRAV} results obtained for the neighbor clump as a proxy for the binary properties expected 
from collapse of the actual clump.  This gives us the approximate SAM of the final binary for the clump in question. The 
procedure is repeated for all clumps produced in the SI simulation and the model SAM distribution is constructed from that.  

\begin{figure}
\epsscale{0.7}
\plotone{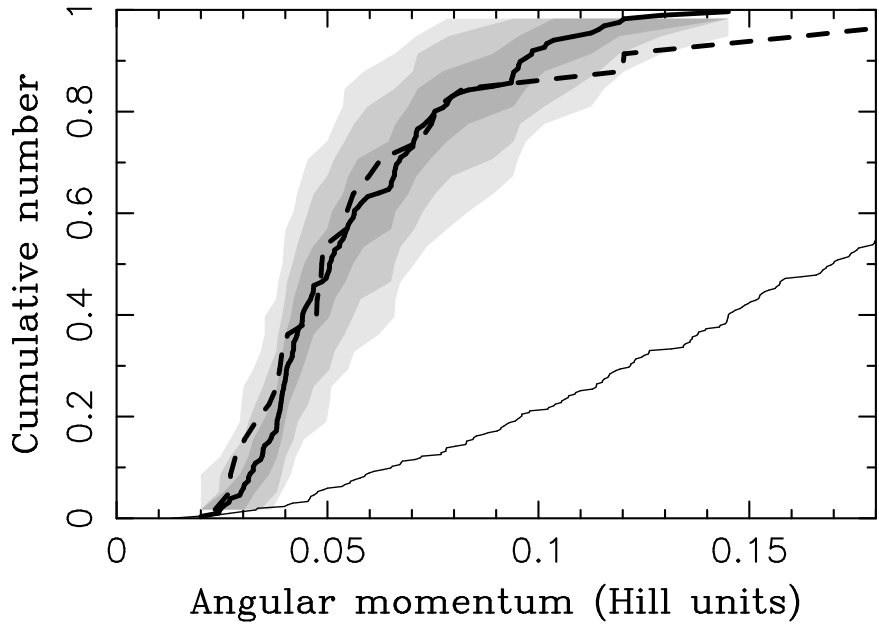}
\caption{The cumulative distribution of the scaled angular momentum for known CC binaries (thick solid line) and model 
binaries obtained for A12 with $t_{\rm A2P} = 38/\Omega$ (dashed line). Here we used $C_{\rm R}=0.5$, $f=1.5$ and $r_{\rm peb}=1$ cm. 
For reference, the thin solid line shows the original SAM distribution of clumps at $t_{\rm A2P} = 38/\Omega$ (Fig. 5). The
shaded areas correspond to the 68.3\%, 95.4\% and 99.7\% confidence intervals (from darker to lighter shade). They were 
determined by sub-sampling the model results to 29 data points -- statistics equal to the number of known CC binaries 
(Table 1). We generated a large number of random trials with the sub-sampled model distributions and determined the uncertainty 
range corresponding to each confidence interval.}    
\label{cum1}
\end{figure}

The model SAM distribution of binaries shows a good agreement with observations (Fig. \ref{cum1}).\footnote{The results 
for A12 and $t_{\rm A2P}=40/\Omega$ are broadly similar but the CDF is wiggly in this case due to the poor sampling of parameter 
space (stars in Fig. \ref{criter}b). It may seem surprising that the distribution of binary SAM is insensitive to $t_{\rm A2P}$
given that the SAM distribution of clumps is narrower for $t=40/\Omega$ than for $t=38/\Omega$ (Fig. \ref{cumul}b). This 
happens because the binary properties depend on $l_{0.5}$, which is more stable over time than $l$, and a larger fraction 
of $l$ is retained in the final binaries for $t=40/\Omega$, because the SI clumps are generally more compact 
at $t=40/\Omega$.} 
The CDF starts at $l_{\rm b}=0.02$-0.03 and steeply rises such that about 80\% of model and observed 
binaries have $l_{\rm b}<0.08$. The distributions differ for $l_{\rm b}>0.08$, because several known binaries have very large 
SAM values: there are four known CC binaries with $l_{\rm b}>0.1$: 2001 QW322, 2000 CF105, 2003 UN284 and 2006 BR286 (Sect. 8.5, 
Table 1). Notably, the 2000 CF105 binary with $l_{\rm b} = 0.145$ has the smallest components of all known CC binaries ($R_1 \simeq 31.5$ 
km and $R_2 \simeq 25$ km). The 2003 UN284 binary is one of the blue binaries identified in Fraser et al.\ (2017). The blue 
binaries presumably formed at $\sim30$-40 au and were pushed out to 42-47 au during Neptune's migration. These properties 
represent interesting clues to the formation of equal-size binaries with $l_{\rm b}>0.1$. 

We conclude that the gravitational collapse model can reasonably well explain the angular momentum distribution of observed 
wide/equal-size binaries (see Appendix B for a discussion of the long-term stability of CC binaries).
It remains to be understood whether it can explain the tight binaries as well. The tight binaries are 
difficult to detect observationally (Noll et al.\ 2008) and they are difficult to obtain in our model because of the resolution 
issues. We thus believe that the SAM distributions in Fig. \ref{cum1} should continue into the realm of 
tight/equal size binaries with $l_{\rm b}<0.02$-0.03. An example of this is the Patroclus-Menoetius binary 
with $l_b\simeq0.018$, a member of the Jupiter Trojan population and one of the targets of the upcoming NASA {\tt Lucy} 
mission (Levison et al.\ 2019). The close relationship of Jupiter Trojans to KBOs (Morbidelli \& Nesvorn\'y 2020) suggests that 
the tight/equal-size binaries should exist in the Kuiper belt as well (Nesvorn\'y et al.\ 2018).
 
\subsection{Collapse dynamics}

\begin{figure}
\epsscale{0.49}
\plotone{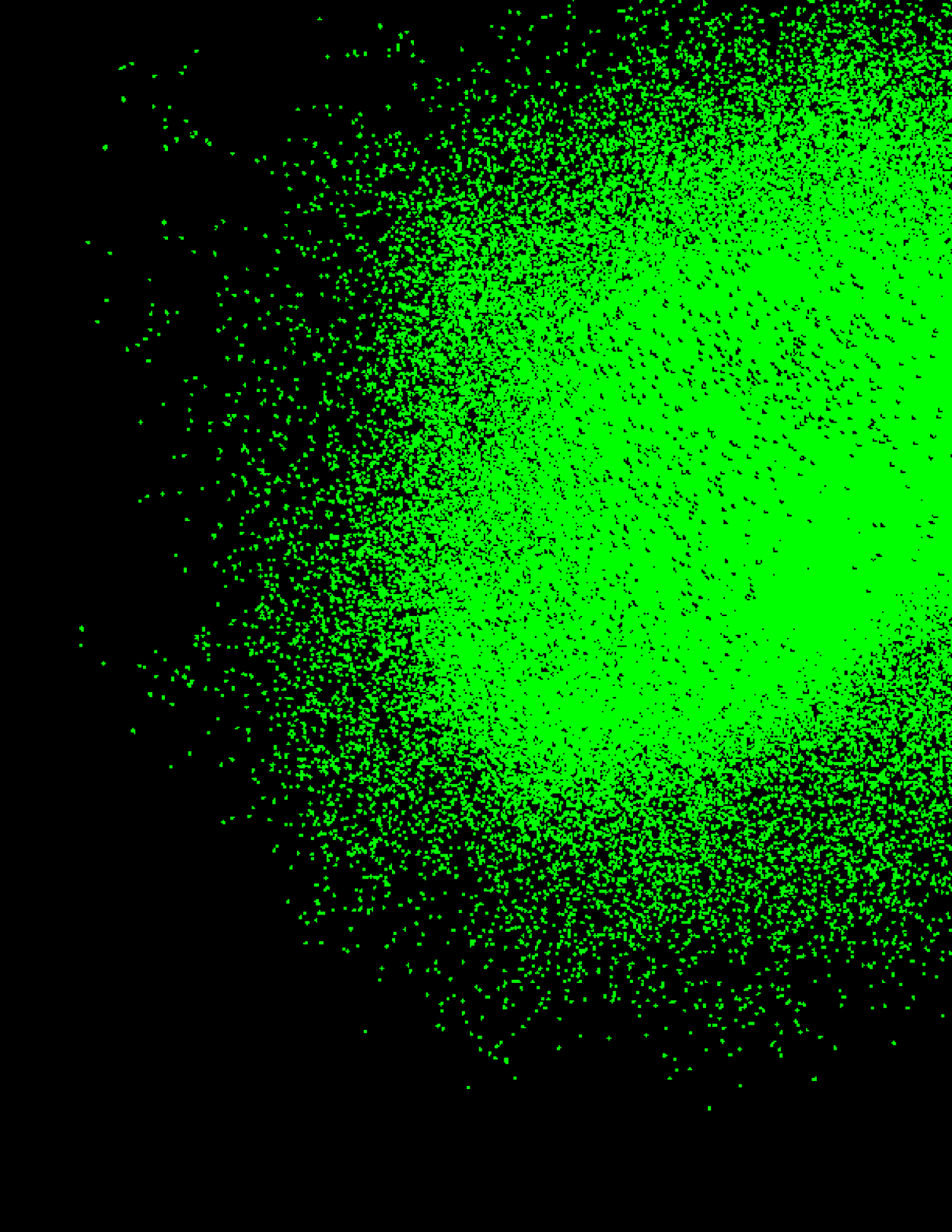}
\plotone{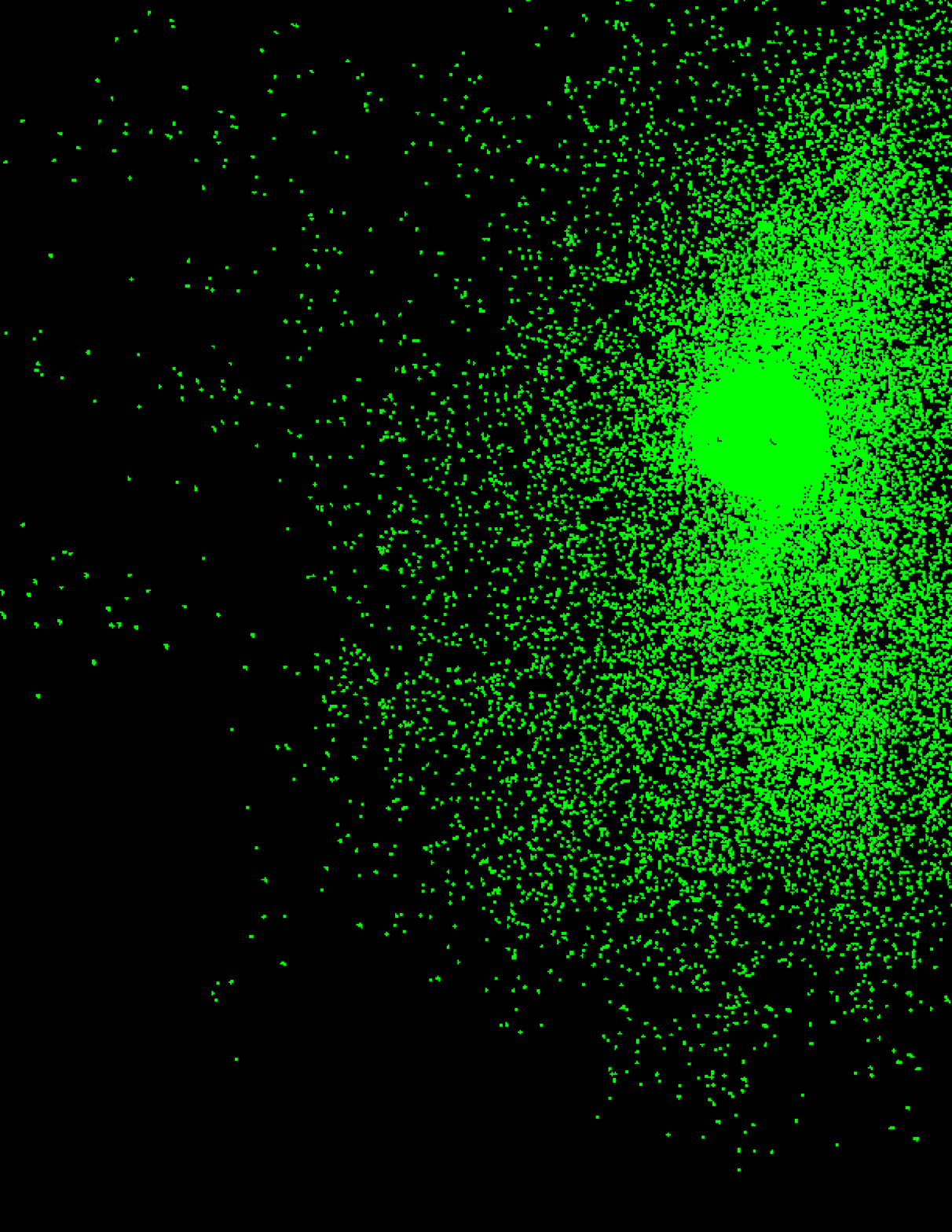}\\[1.mm]
\plotone{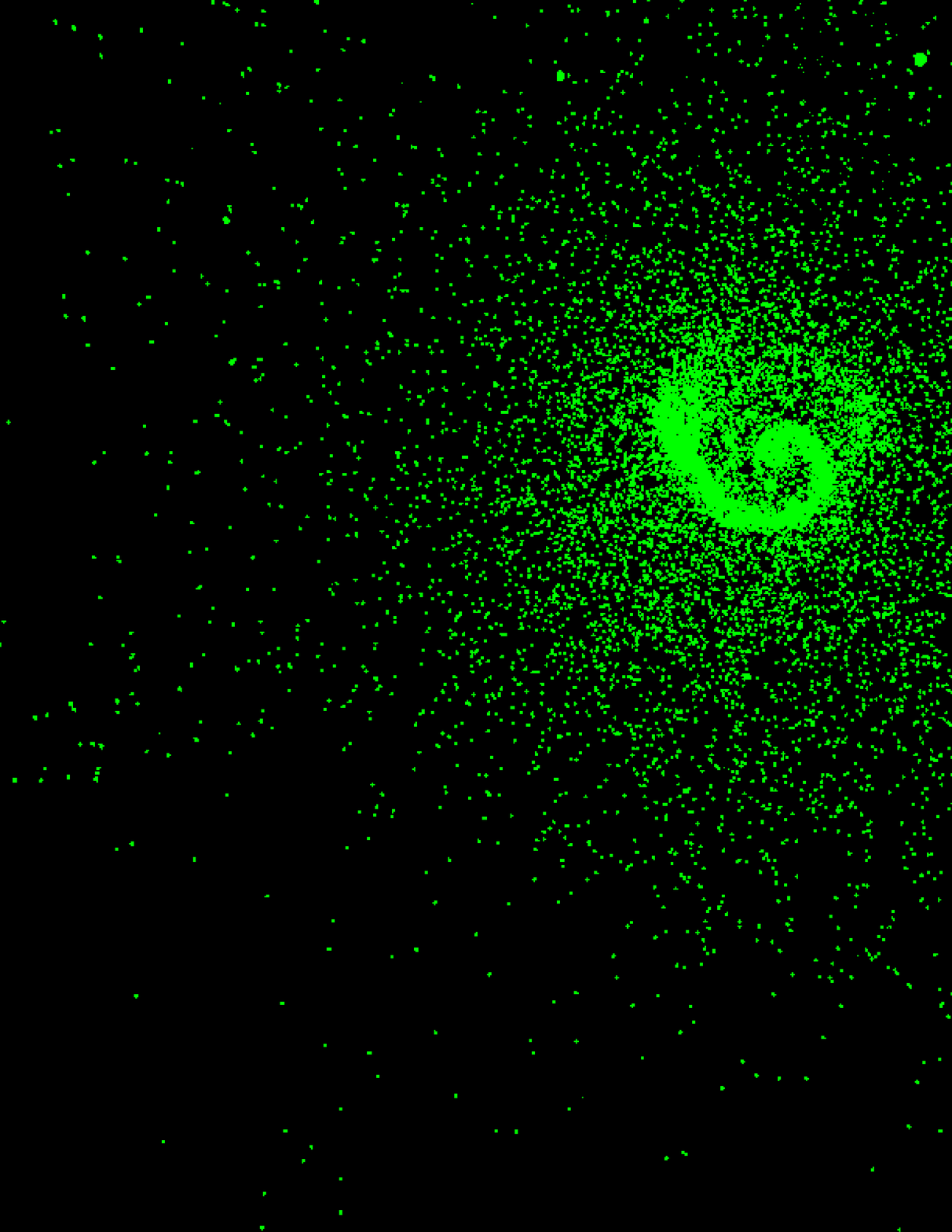}
\plotone{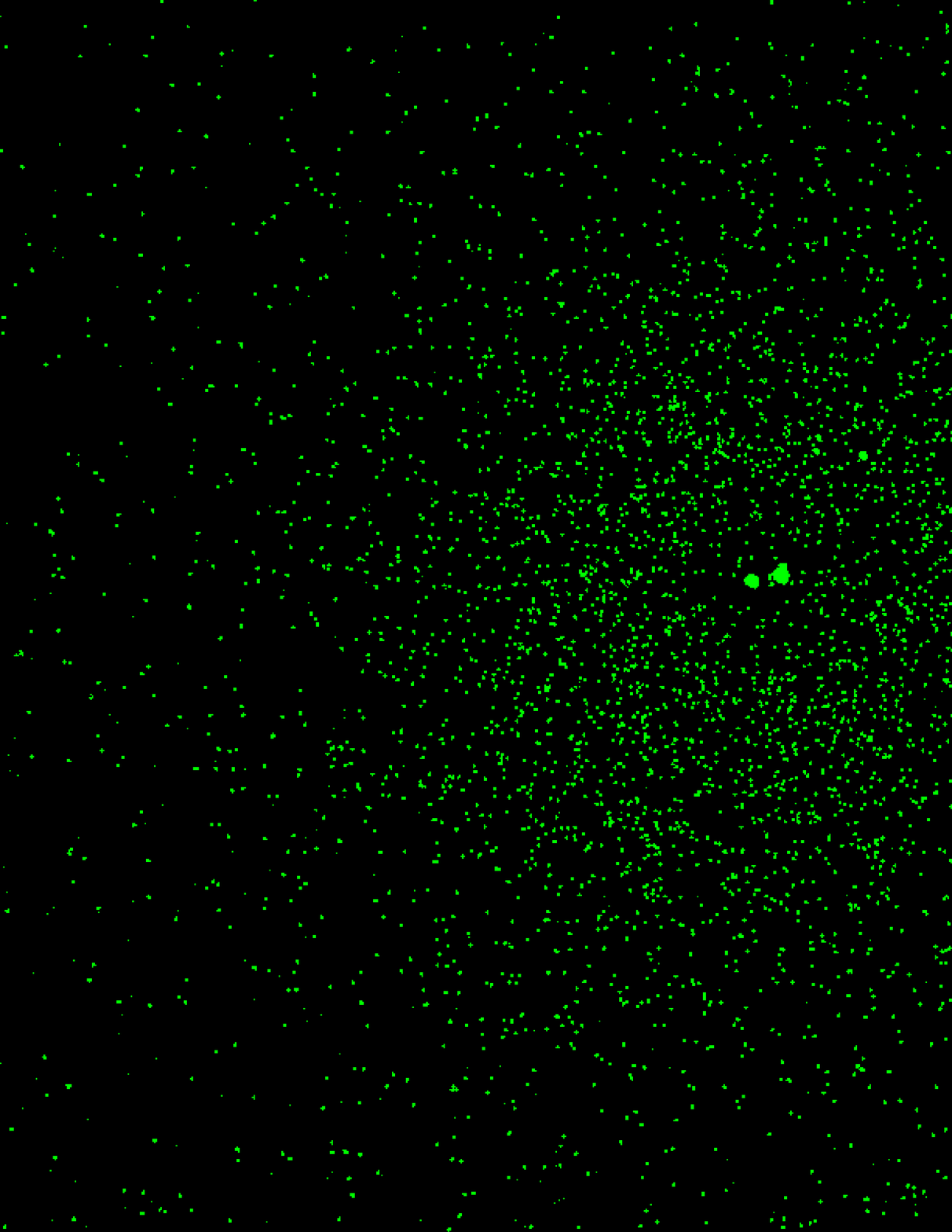}
\caption{Four snapshots from our {\tt PKDGRAV} collapse simulations of clump a00 ($t_{\rm A2P}=40/\Omega$, $C_{\rm R}=0.5$,
$f=1.5$, $r_{\rm peb}=1$ cm): $t=0$ (top left), 6 (top right), 13 (bottom left) and 50 yr (bottom right). 
The view is projected down the clump angular momentum axis. Each frame is 
450,000 km across. Note the pebble disk in the top-right panel and its subsequent disruption in the bottom-left panel. 
The equal-size binary in the bottom-right panel has $R_1=55.6$ km, $R_2=38.8$ km, $a_{\rm b}=10400$ km, $e_{\rm b}=0.19$ 
and $l_{\rm b}=0.054$.}    
\label{frame1}
\end{figure}

Collapsing pebble clumps share similar characteristics. During the first stage, lasting $<10$ yr, each cloud 
collapses toward its center of mass (Fig. \ref{frame1}). The inner parts of clouds are typically more dense and collapse 
faster, leading to a formation of a disk structure in the center (top-right panel of Fig. \ref{frame1}). The disk radius 
is proportional to the initial SAM value (or, more accurately, to $l_{0.5}$), with the largest disks being produced for clumps 
with the largest initial SAM. The detailed structure of the disk (e.g., its thickness) may be affected by the dense packing of 
SPs and higher-resolution runs will be needed to test convergence. The disk is short-lived, either because it is inherently
unstable or because of the infall of material with the high angular momentum, and is dispersed (bottom-left panel of 
Fig.~\ref{frame1}). Eventually, an equal-size binary emerges from the dispersed disk (bottom-right panel of Fig. \ref{frame1}), 
gradually consuming nearly all the disk mass. This last stage is completed in tens of years after the initial collapse.      

\begin{figure}
\epsscale{0.49}
\plotone{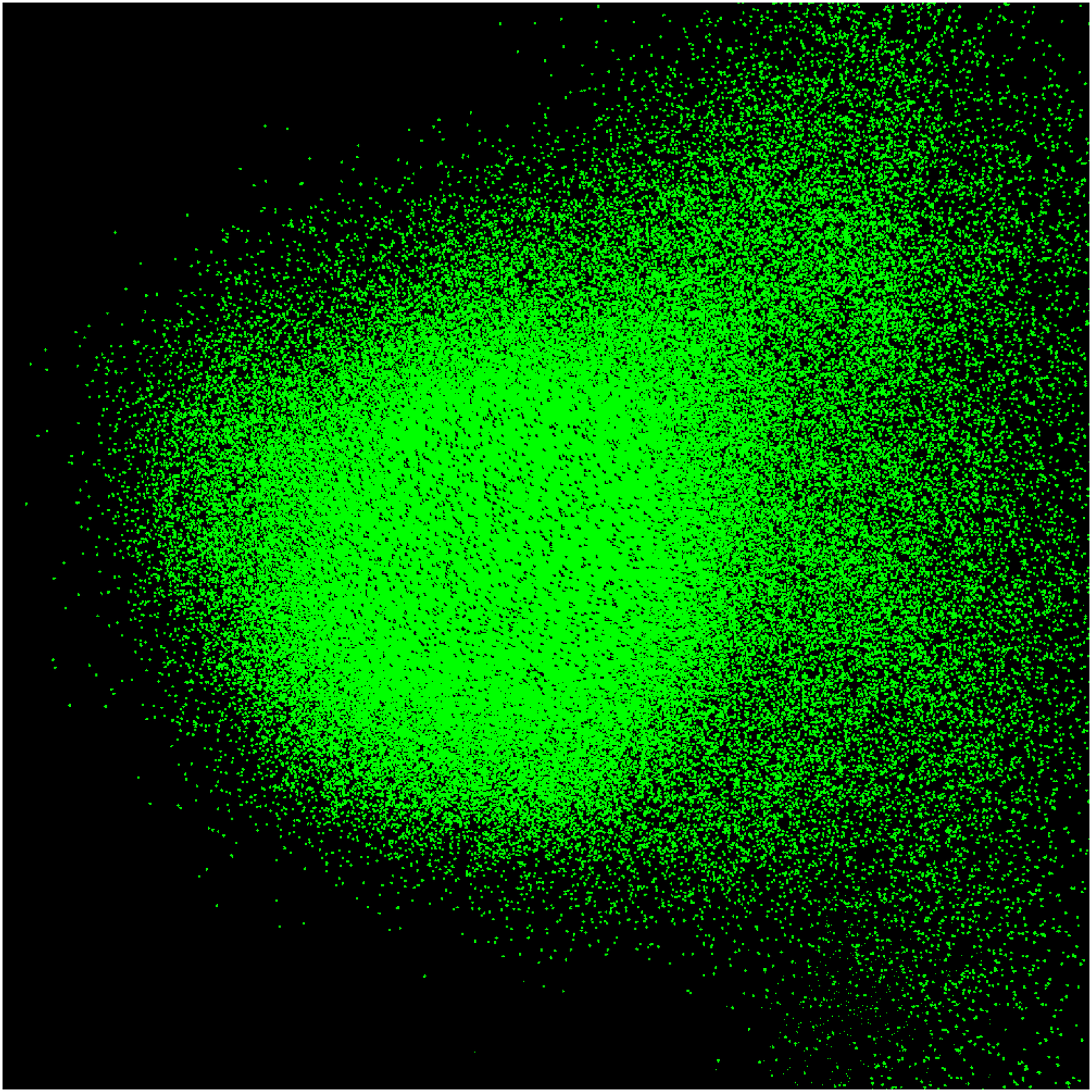}
\plotone{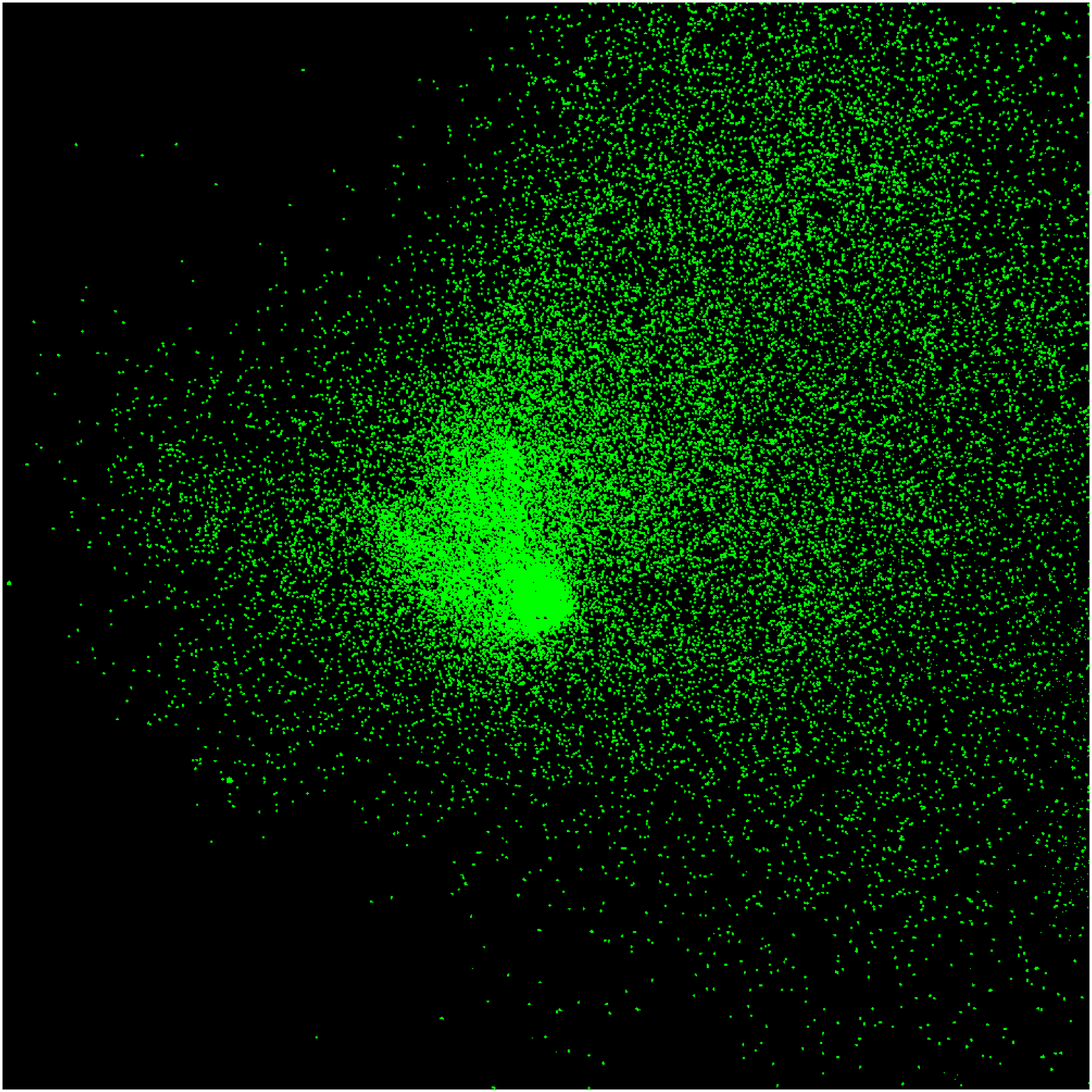}\\[1.mm]
\plotone{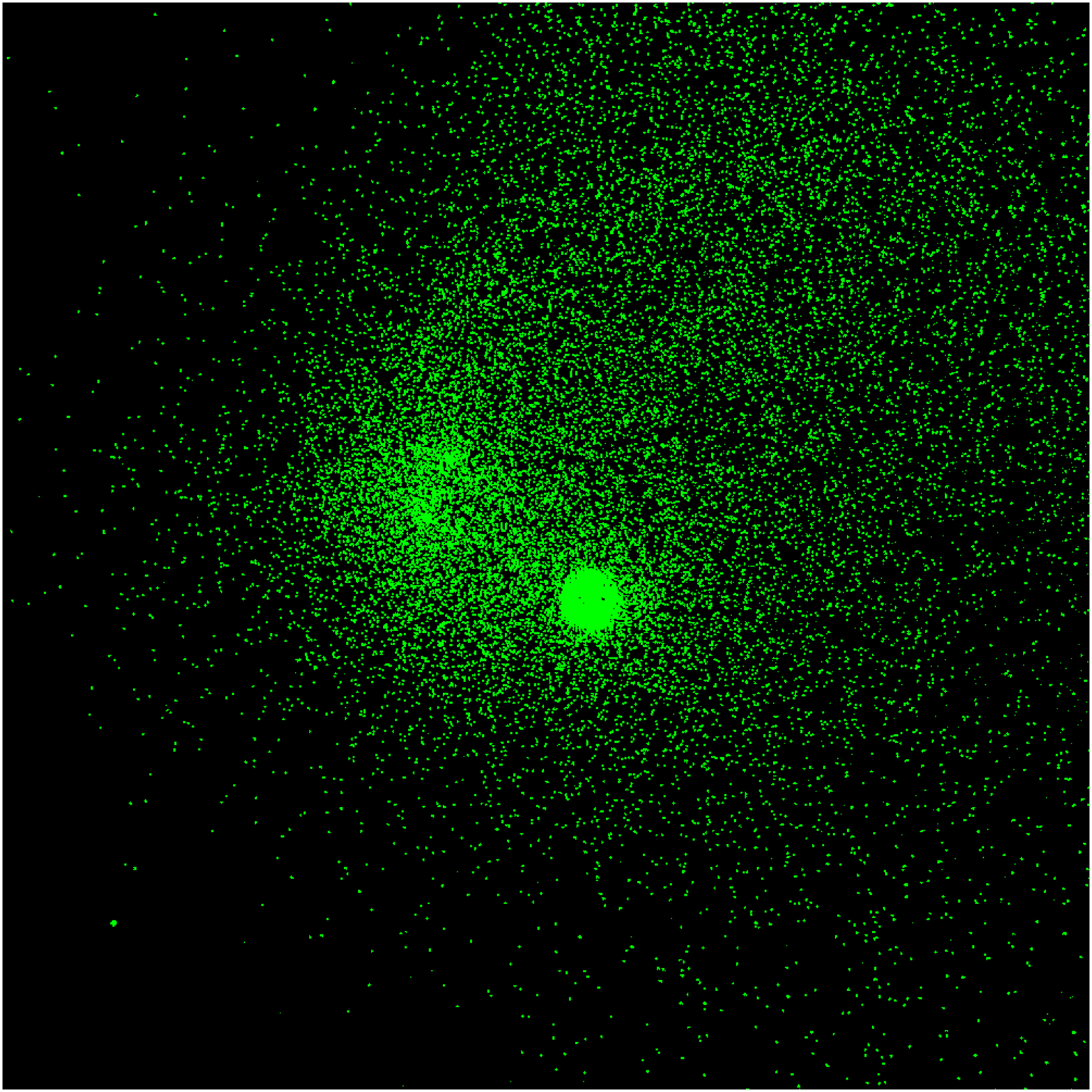}
\plotone{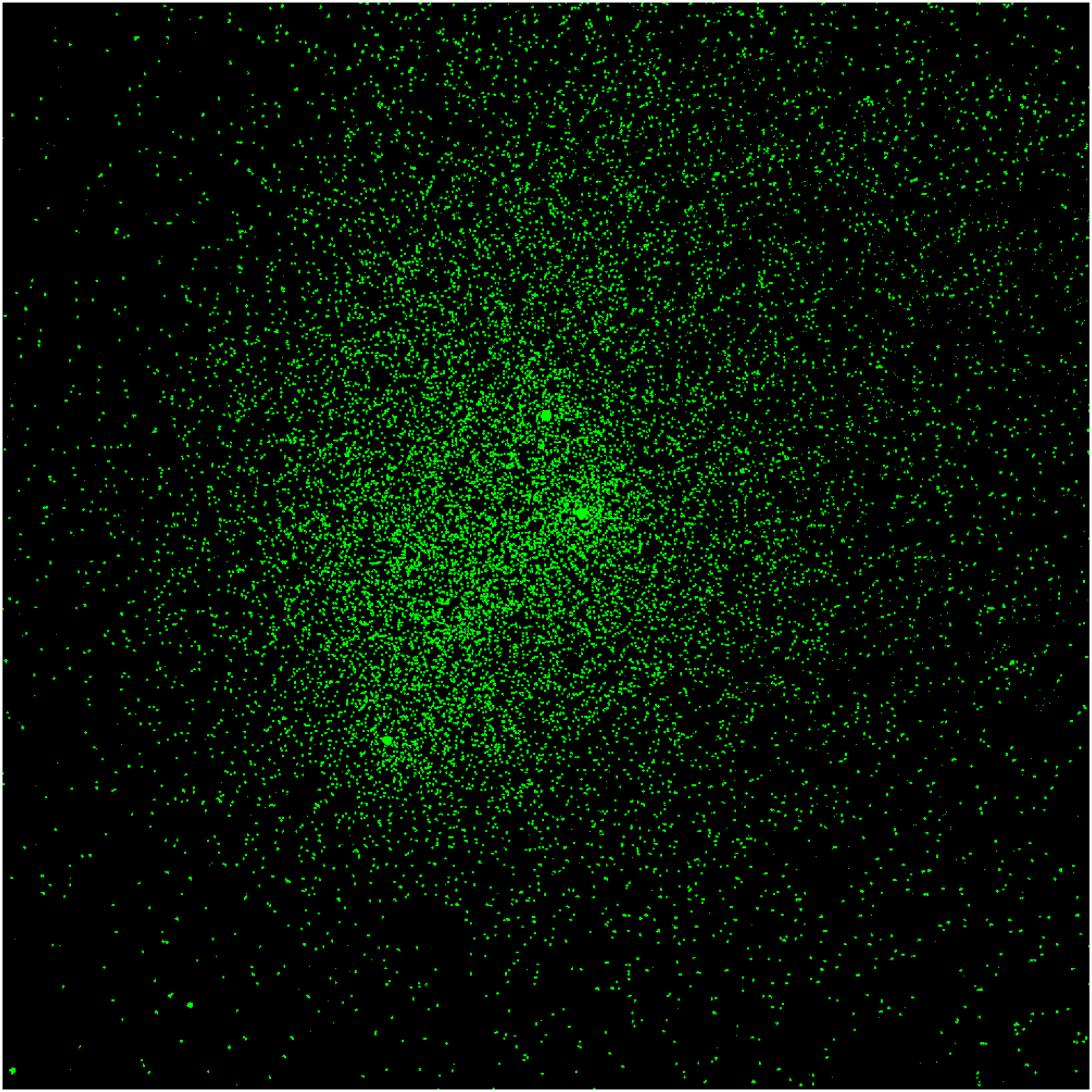}
\caption{Four snapshots from our {\tt PKDGRAV} collapse simulations of clump a03 ($t_{\rm A2P}=40/\Omega$, $C_{\rm R}=0.5$,
$f=1.5$, $r_{\rm peb}=1$ cm): $t=0$ (top left), 10 (top right), 14 (bottom left) and 20 yr (bottom right). 
The view is projected down the clump angular momentum axis. Each frame is 
450,000 km across. The pebble disk that can seen in the top-right and bottom left panels collapses into a single 
planetesimal by $t=20$ yr. Two other planetesimals form from local concentrations of pebbles outside the disk. 
Finally, the three massive planetesimals undergo an exchange reaction and one of them is ejected. A wide/equal-size 
binary is left behind with $R_1=50.5$ km, $R_2=37.2$ km, $a_{\rm b}=32000$ km, $e_{\rm b}=0.28$ and $l_{\rm b}=0.104$.}    
\label{frame2}
\end{figure}

The formation of solitary planetesimals in our simulations follows the same pattern, except that the disks produced by 
low-SAM pebble clumps are tiny and the inflated SPs come into contact and merge. Higher-resolution simulations will be needed to 
better resolve the compact disks and study the formation of tight/contact binaries. The high-SAM clumps show more complicated dynamics 
with several accretion centers leading to the accretion of several large planetesimals. The gravitational interaction of these bodies 
is complicated and often leads to mass ejection. An example of this is shown in Fig. \ref{frame2}, where three planetesimals 
form by $t=14$ yr (bottom-left panel), and one of them is ejected shortly after that. In this case, the final equal-size 
binary is widely separated and has a relatively large SAM content ($l_{\rm b}=0.104$). In other cases, a hierarchical multiple 
system with roughly coplanar and non-crossing orbits can emerge.

\subsection{Runs B22 and C203}

A similar analysis was performed for runs B22 and C203 (Table 4). In total, we completed 16 {\tt PKDGRAV} simulations for 
clumps extracted from B22 ($t_{\rm A2P}=41/\Omega$) and 8 simulations for clumps from C203 ($t_{\rm A2P}=117.6/\Omega$) 
Unlike for A12, here we only used one random seed to generate the initial distribution for each clump; the statistical 
variability of these results is therefore expected to be larger. Ten out of 16 (63\%) clumps from B22 and 6 out of 8 
(75\%) clumps from C203 produced equal-size binaries with $R_2/R_1>0.5$. The properties of model binaries are similar to the ones 
reported from A12: (1) the radius ratio reaches up to $R_2/R_1=0.96$ (clump b15), (2) the SAM values of 
equal-size binaries range between $l_{\rm b}=0.03$ and 0.15, (3) the binary eccentricities are nearly uniformly distributed 
in the $0<e_{\rm b}<0.8$ interval, and (4) the binary inclinations with respect to the initial angular momentum vector are 
generally small (e.g., 71\% from B22 have $i_{\rm b}<20^\circ$ and 100\% from B22 have $i_{\rm b}<35^\circ$; the more 
misaligned cases are more unequal). Due to the resolution issues discussed in Sect. 8.6, we do not obtain binaries with 
$R_2/R_1>0.8$ and $l<0.05$, and the model binaries do not populate orbits with $q_{\rm b}<35(R_1+R_2)$. 

The binary formation criterion (Sect. 8.6) works pretty well for B22 and C203 (Fig. \ref{criter2}). Eight out of 9 
clumps selected from B22 with $l_{0.5}>0.07$ produce equal-size binaries. The only exception to this rule is clump 
b03 with $l_{0.5}=0.097$. The initial clump structure shows two concentration centers and is reminiscent to that of 
clump a01 at $t=41/\Omega$. By $t=100$ yr, the b03 clump gives birth to a system with four moons having $R_j/R_1=0.22$
-0.35. Some of these moons have crossing orbits and we thus prolonged to simulation to see how the system evolves.
Only one moon with $R_2/R_1=0.36$ survived by $t=1000$ yr. The other moons were accreted on the primary or ejected 
from the system. From six B22 clumps having $l<0.07$ initially, four produced unequal size binaries and one ended 
with a single planetesimal. The only exception is clump b04 with $l_{0.5}=0.056$, which produced an equal size
binary with $R_2/R_1=0.67$. We have simulated this clump with a different random seed and found that the additional 
simulation produced a binary with $R_2/R_1=0.51$. More work will be needed to fully characterize the statistical 
variability of the results.   

\begin{figure}
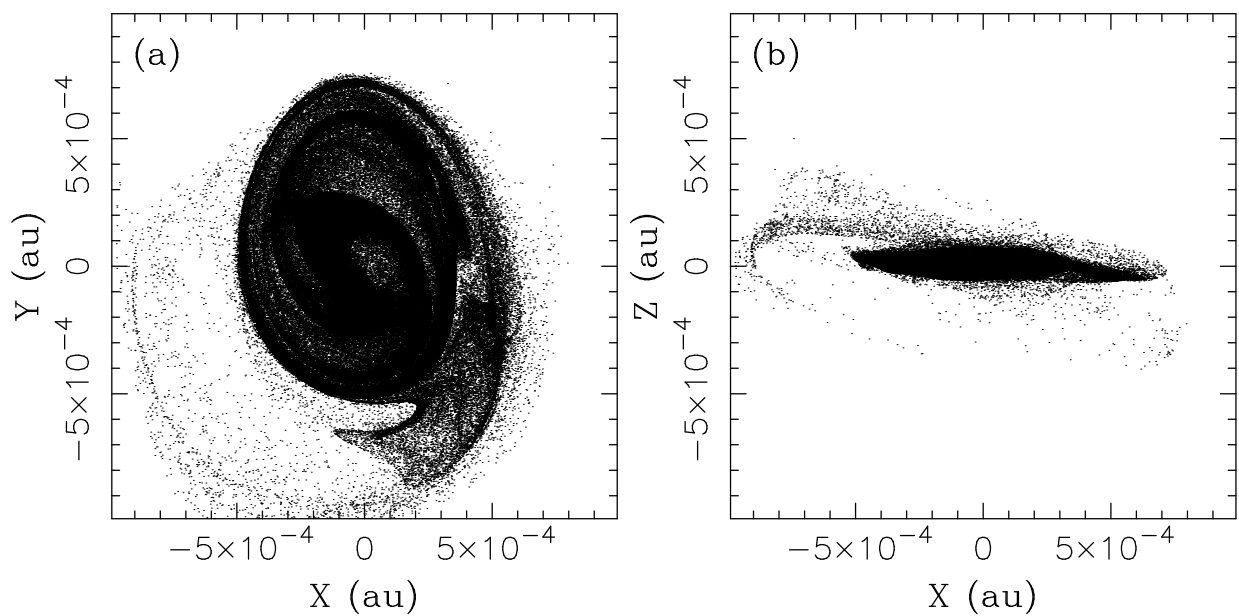

\epsscale{0.49}
\plotone{fig17a.eps}
\plotone{fig17b.eps} 
\caption{The structure of the c05 clump at $t=117.6/\Omega$. The angular momentum vector points along the $Z$ axis. 
The total mass of the clump corresponds to a sphere with equivalent radius $R_{\rm eq} =136$ km (for a reference density 
$\rho=1$ g cm$^{-3}$). The boxes shown here are roughly 300,000~km across. For scale, the Hill radius is $R_{\rm H} \sim 
815,000$ km (for $a \simeq 45$ au).}
\label{c05}
\end{figure}

The results for C203 are similar (Fig. \ref{criter2}b) with 7 out of 8 simulated clumps (88\%) satisfying the criterion
(5 have $l_{0.5}>0.07$ and form equal size-binaries, 2 have $l_{0.5}<0.07$ and form unequal-size binaries). 
 The c05 clump is an interesting exception. This clump initially has $l \simeq l_{0.5}=0.057$ and 
yet it forms an equal-size binary with $R_2/R_1=0.59$ (Table 4). It is one of the largest clumps selected from C203, 
has low initial obliquity and is initially very flat (Fig. \ref{c05}). The two binary components form in the overdense 
regions near where in the inner spiral arms bend (Fig. \ref{c05}a). We have not investigated in detail how often the 
initial clumps are flattened but we have not seen this happening again in $\sim$20 other cases we looked at. This 
suggests that cases such as c05 are uncommon and their relevance for the planetesimal formation is relatively low. 

The cumulative distributions of $l_{\rm b}$ obtained for runs B22 and C203 fit observations reasonably well but they are
more choppy than the one shown in Fig. \ref{cumul}. This is most likely related of the approximate nature of the method 
used here to construct the model distributions (Sect. 8.7). To make the predictions more accurate we will need to: (1) 
increase the resolution of the {\tt PKDGRAV} simulations, and (2) simulate collapse of a larger number of selected 
clumps. For example, it is conceivable that {\it all} clumps from a given {\tt ATHENA} run can be simulated with {\tt PKDGRAV}. 
Increasing the resolution will require a larger computer power and some fix to bypass the computational bottleneck during the disk 
stage where packed SPs often collide (this stage is not efficiently parallelized in the hard sphere flavor of {\tt PKDGRAV}).
If the predictions can be made more accurate, it should eventually be possible, by comparing them with observations of the CC 
binaries, to constrain the SI parameters that applied to the formation of planetesimals in the outer solar system.
  
\section{Conclusions}

The main conclusions of this work can be summarized as follows:
\begin{enumerate}
\item The streaming instability simulations produce pebble clumps with vigorous rotation that are conducive for the formation
of equal-size planetesimal binaries. 
\item We simulate gravitational collapse of pebble clumps and find it involves several stages: (i) the initial infall of dense 
inner regions leads to a disk formation, (ii) the binary system accretes from the disk material, (iii) the disk is disrupted by 
instabilities and/or continued infall of material, and (iv) a binary forms, ejects mass, and evolves to its final configuration.
\item The binary properties depend on the initial value of 
the scaled angular momentum (SAM or $l$; Sect.\ 6) and its radial distribution. Low-$l$ clumps produce close (or contact) 
binaries and single planetesimals. Compact high-$l$ clumps with $l>0.07$ for $r<0.5 R_{\rm H}$ give birth to wide equal-size 
binaries ($R_2/R_1>0.5$ and $a_{\rm b}/R_{\rm b}>35$, where $R_{\rm b}$ is the sum of binary component radii, $R_{\rm b}=R_1+R_2$). 
High-$l$ clumps with a flat radial distribution produce hierarchical multiple systems with large moons ($R_j/R_1=0.1$-0.5).
\item We find that 10-75\% of the original SAM and 50-100\% of the original mass of a pebble clump is converted to the final 
binary system. The remaining $<50$\% of the original mass remains in pebbles and small planetesimals. The mass loss should 
be taken into account when estimating the initial mass function of planetesimals (e.g., Simon et al.\ 2017).
\item SI-triggered gravitational collapse can explain the angular momentum distribution (or, equivalently, the mutual 
separations) of the equal-size binaries found in the Kuiper belt (Sect 8.7; Grundy et al.\ 2019, Noll et al.\ 2020).
\item The tilt of the orbital plane of binaries with respect to the initial angular momentum vector of clumps is generally small 
($\lesssim 20^\circ$). This justifies the assumption of Nesvorn\'y et al.\ (2019) that the obliquity distribution of clumps can
be used to predict the inclination distribution of binaries. 
\item The resolution of {\tt PKDGRAV} simulations will need to be increased to $\gtrsim 10^6$ superparticles to be able to 
resolve the tight (and contact) binaries (mutual pericenter $<35 R_{\rm b}$). The high-resolution collapse simulations tend to 
produce binary components that are more similar in size compared to the low-resolution runs.
\item Roughly 10\% of our collapse simulations produce hierarchical systems with two or more large moons. These systems should 
be found in the Kuiper belt when observations reach the threshold sensitivity. 
\item The changes of binary properties produced by the collisional/dynamical evolution of Cold Classicals over the age of the 
solar system are relatively minor (Appendix B; also see Porter \& Grundy 2012).    
\end{enumerate}


\acknowledgements
The work of DN was funded by the NASA EW program. RL acknowledges support from NASA headquarters under the NASA Earth 
and Space Science Fellowship Program grant NNX16AP53H.
ANY acknowledges support from NASA Astrophysics Theory Grant NNX17AK59G and from 
NSF grant AST-1616929. DCR acknowledges support from NASA Solar System Workings grant NNX15AH90G.
RM acknowledges the support of the Swiss National Science Foundation (SNSF) through the grant P2BEP2\_184482.
 
\appendix

\section{Collisions vs.\ Drag at Handoff}
%
The transition from {\tt ATHENA} simulations of SI to {\tt PKDGRAV} simulations of collisional collapse involve a discontinuous 
transition in the included physics.  In  {\tt ATHENA}, hydrodynamics and gas drag are included and collisions neglected. For 
{\tt PKDGRAV} simulations the situation is reversed.  This transition is motivated because at high particle densities, particle 
collisions are more important than drag forces, as a dissipation mechanism.  NYR10 computed analytic estimates of the relevant 
ratios of particle collision time to aerodynamic stopping time.  We review and update these estimates below.

For the current {\tt PKDGRAV} simulations we can directly measure the collision rate, instead of using simpler estimates.   
To understand the collisional state of a clump at the handoff we considered clump b00 at $\Omega t = 39.5$ from  {\tt ATHENA} output.  
This particle distribution was transferred to  {\tt PKDGRAV} and simulated for a brief time, $0.1$ yr $ \simeq 2 \times 10^{-3}/\Omega$ 
(at 45 AU).  The number of collisions measured imply a per particle collision rate of $t_{\rm coll} = 0.065/\Omega$.  This normalized 
timescale is shorter than the normalized stopping times $\tau_s = 0.3$-$2$ in the {\tt ATHENA} simulations.  Thus the neglect of 
drag in the  {\tt PKDGRAV} simulations is justified, especially inside $0.3 \RH$ where most particles are already concentrated.  
This analysis suggests that collisions should be included in the later stages of planetesimal formation in the SI simulations, 
perhaps using an approximate treatment of collisions as in \citet{johansen12}, using {\tt PENCIL}.  Adding collisional effects to 
{\tt ATHENA}  (and/or analyzing {\tt PENCIL} simulations with {\tt PLAN}, \citealp{li19}) is left to future work.

Analytic estimates of $t_{\rm coll}/t_{\rm stop}$ are useful for a more general understanding of whether collisional or drag dissipation 
should be more important (NYR10, \citealp{klahr20}).  We redo the calculation of NYR10 to more carefully retain numerical prefactors, 
because even though the estimate is highly idealized and precision is not physically warranted, it allows a uniform comparison with 
detailed calculations, as presented above, to tune coefficients.  We assume a spherical cluster of particles of total mass $M$ and 
individual particle masses $m = (4 \pi/3) \rho_s s^3$ and mutual cross section $\sigma = 4 \pi s^2$.  The particles uniformly fill 
a sphere of radius  $f_H R_H$, i.e. a fraction $f_H < 1$ of the Hill radius.  Note that a uniform density ignores the central 
concentration of any realistic cluster, which a smaller $f_H$ can approximately account for.  The relative speed of particles is 
assumed to be $v = f_v \sqrt{G M/(f_H R_H)}$ with another correction factor $f_v$.  (Note that a self-consistent velocity from the 
virial theorem is not possible for a constant density sphere, which has an infinity polytropic index, $\gamma$, while elastically 
colliding ``monatomic" particles should have $\gamma = 5/3$.)  The ratio of $t_{\rm coll} = (n \sigma v)^{-1}$ to $\Omega t_{\rm stop} = 
\pi \rho_s s/(2 \Sigma_g)$ is thus
\begin{align}
\frac{t_{\rm coll}}{t_{\rm stop}} &= \frac{8}{27  \cdot 3^{1/6}} \frac{f_H^{7/2}}{f_v}  \frac{\Sigma_g a^2}{M^{1/3} M_\odot^{2/3}} \approx 3.2 
\frac{f_H^{7/2}}{f_v} \sqrt{\frac{a}{45\; {\rm AU}}} \frac{100\; {\rm km}}{R_{\rm eq}} \, .
\end{align} 
The final numerical estimate uses the MMSN of \citet{cy10} and an internal density of $1$ g cm$^{-3}$ to compute the equivalent radius 
of the collapsed cluster.  With strong central concentration ($f_H \lesssim 0.5$) the collapse stage should be collision dominated 
in the outer disk, as our more careful estimate above confirms.  If we fix $t_{\rm coll}/t_{\rm stop} = 0.065/2$ and $f_H = 0.3$ to 
match the conditions of clump b00 above, then the remaining ``fudge" factor $f_v = 1.46 (100~{\rm km}/R_{\rm eq})$.  While physically 
a small $f_v < 1$ might be expected for a clump with rotation being larger than random motions, the agreement of simulations with 
the approximation is good.

\section{Binary survival}

We briefly discuss additional effects that may have affected KBO binaries. The goal is to understand whether
the observed properties of CC binaries are truly primordial or whether they may have been altered during the subsequent evolution.  
A binary orbit can be affected by gravitational encounters with massive perturbers (Nesvorn\'y \& Vokrouhlick\'y 2019). However, 
this should not be a major issue for the CC binaries, because CCs did not engage in encounters with Neptune (Parker \& Kavelaars 2010). 
The CC binaries could potentially be affected by encounters with planetary-size bodies, if these bodies formed in the original 
trans-Neptunian disk and were scattered by Neptune outward (they would temporarily overlap with the CC population before being 
ejected from the solar system; Shannon \& Dawson 2018). 

We now consider the collisional survival. The mutual orbit of a binary can be affected by small impacts into its components. The 
binary may become unbound if the velocity change imparted by an impact exceeds the binary's orbital speed (typically meters per second for 
known KBO binaries; Petit \& Mousis 2004, Petit et al. 2008). We investigate this process with the collision code that we previously developed 
(Morbidelli et al.\ 2009, Nesvorn\'y et al.\ 2011). The code, known as {\tt Boulder}, employs a statistical method to track the collisional 
fragmentation of planetesimal populations. A full description of the {\tt Boulder} code, tests, and various applications can be 
found in Morbidelli et al.\ (2009), Levison et al.\ (2009) and Bottke et al.\ (2010). The binary module in {\tt Boulder}  accounts 
for small, non-disruptive impacts on binary components, and computes the binary orbit change depending on the linear momentum of 
impactors (Nesvorn\'y et al.\ 2011). 

We account for impacts over the whole solar system history. The dynamical model of Nesvorn\'y \& Vokrouhlick\'y (2016) is used
to follow Neptune's migration into the outer planetesimal disk. The disk is assumed to start dynamically cold ($e \simeq 0$ and
$i \simeq 0$) and becomes excited, on a timescale of $\sim$10-30 Myr, by a migrating Neptune. The \"Opik algorithm (Wetherill 1967, 
Greenberg 1982) is used to compute the collision probabilities $P_{\rm col}$ and velocities $v_{\rm col}$ as a function of time. 
We monitor $P_{\rm col}$ and $v_{\rm col}$ among CCs and between CCs and other planetesimals included in the dynamical simulation. 
The average values of these parameters are computed by taking the mean over all pairs of bodies at each time. The changing conditions 
are implemented in the {\tt Boulder} code, which is then used to determine the collisional survival of CC binaries (see Nesvorn\'y 
\& Vokrouhlick\'y 2019 for details). 

\begin{figure}
\epsscale{0.5}
\plotone{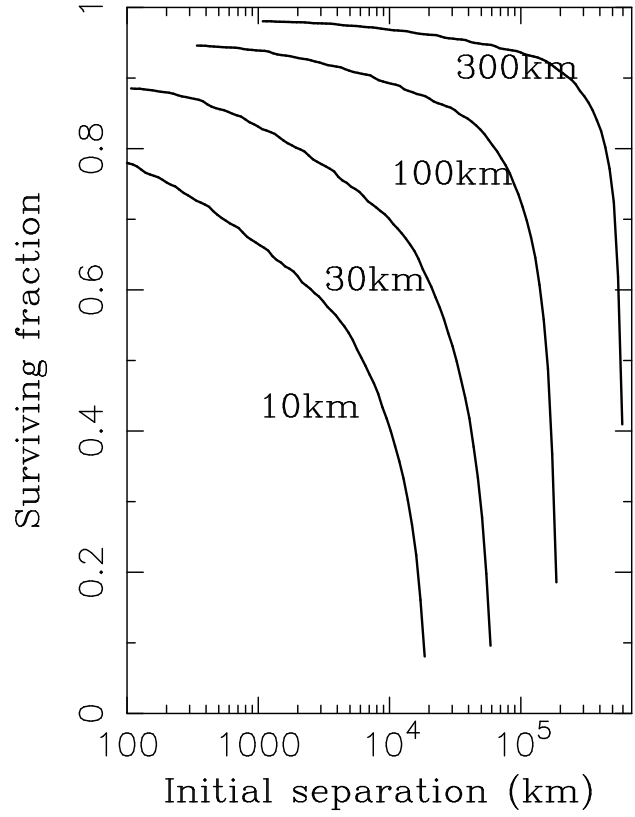}
\caption{The collisional survival of equal-size binaries in the CC population. The plot shows the surviving fraction 
for $R_1+R_2$=10, 30, 100 and 300 km as a function of initial binary separation $a_{\rm b}$.}   
\label{survival}
\end{figure}

We find that the survival chances of CC binaries are generally good but drop when the binary separation approaches 0.5 $R_{\rm H}$
(Fig. \ref{survival}), which is expected because binaries with $a_{\rm b}>0.5$ $R_{\rm H}$ are dynamically unstable (e.g., Porter \& 
Grundy 2012). Most 100-km class CC binaries survive. This contrasts with the low survival chances of binaries in the hot 
population, which are dynamically removed during their implantation into the Kuiper belt (Nesvorn\'y \& Vokrouhlick\'y 2019). 

\begin{figure}
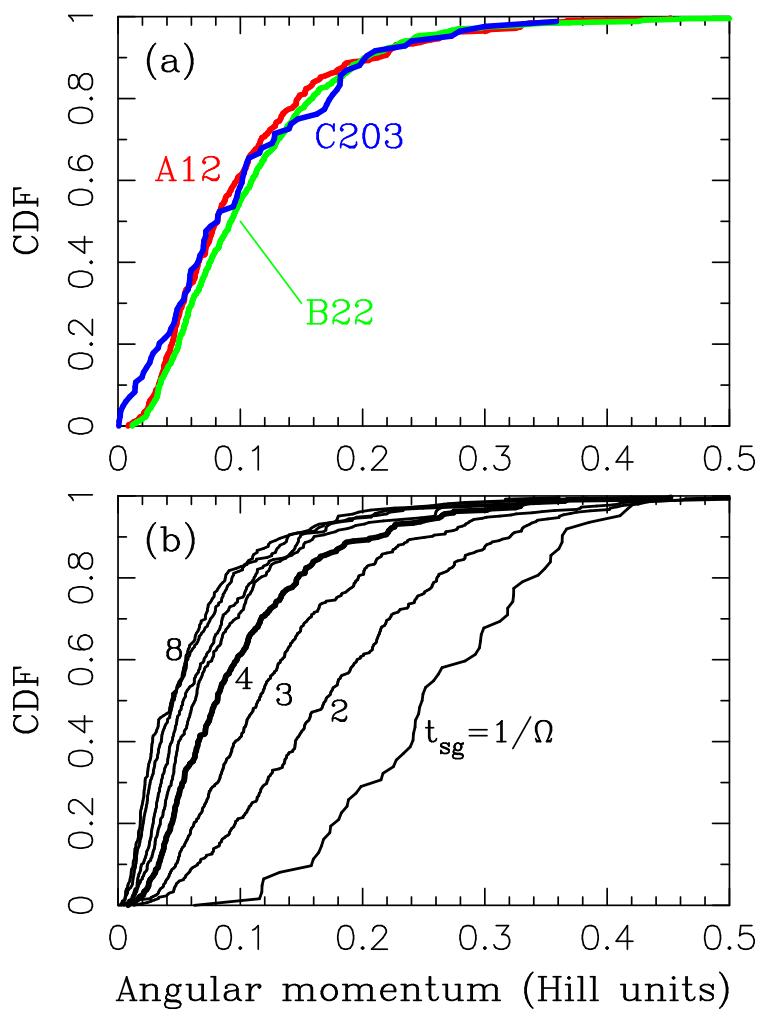

\epsscale{0.6}
\plotone{fig5a.eps}\\[2.mm]
\plotone{fig5b.eps}
\caption{Panel a: The cumulative distribution function (CDF) of SAM in runs A12 (red line), B22 (green) and C203 (blue). 
The clumps were extracted at $t_{\rm sg}=4 / \Omega$ in A12, $t_{\rm sg}=4 / \Omega$ in B22 and $t_{\rm sg}=7.6/\Omega$ in C203.
Panel b: Changing CDF($l$) with time in the A12 simulation.}  
\label{cumul}\end{figure}

\begin{figure}
\epsscale{0.7}
\plotone{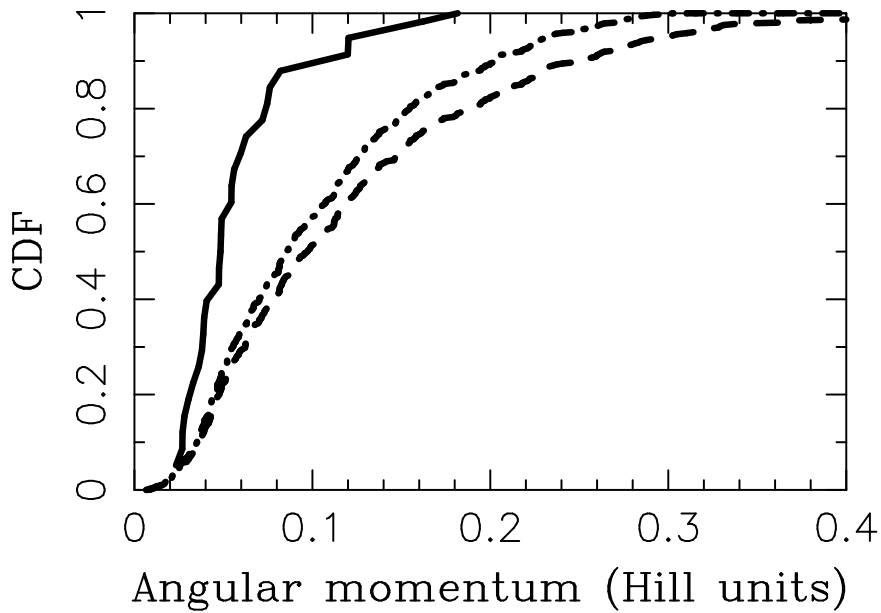}
\caption{The effect of collisional survival on the SAM distribution of CC binaries. The plot shows the SAM for CC binaries 
(solid) and clumps from the A12 run (dashed; $t=40/\Omega$). The dot-dashed line accounts for 
the collisional removal of binaries. For that, we assumed that every SI cloud produces a binary with no mass or angular momentum loss, 
and used the results for $R_1+R_2=100$ km from Fig. \ref{survival} to estimate how the distribution would change if binaries are 
removed.}   
\label{survival2}
\end{figure}

Figure \ref{survival2} shows the SAM distribution for the CC binaries. For that we adopted the SAM distribution of A12 clumps and
$t_{\rm A2P}=40/\Omega$ (bold line in Fig. \ref{cumul}b), and assumed that every SI cloud produces a binary with no mass or angular 
momentum loss. We then used the results from Fig. \ref{survival} to compute how the collisional removal would effect the SAM distribution. 
This turns out not to be a large factor (Fig. \ref{survival2}), because the initial and final distributions are similar. 
We conclude that collisional removal cannot explain the difference between the angular momentum distributions of known binaries 
and the SI clumps. Instead, the difference is related to the angular momentum loss during gravitational collapse and binary 
formation (Sect.\ 8.7).

\clearpage

\begin{table}
\centering
{
\begin{tabular}{lrrrrrr}
\hline \hline
 number & name & prov. desig. & $R_2/R_1$ & $a_{\rm b}/R_{\rm b}$ & $e_{\rm b}$ & $l_{\rm b}$ \\  
\hline
58534   & Logos       & 1997 CQ29   &  0.77 &  113  &   0.546 &   0.053 \\
66652   & Borasisi    &	1999 RZ253  &  0.82 &  39.6 &   0.470 &   0.036 \\
79360   & Sila-Nunam  & 1997 CS29   &  0.95 &  11.5 &   0.026 &   0.024 \\
88611   & Teharonhiawako & 2001 QT297 & 0.73 & 180 &    0.249 &   0.073 \\
123509  & -           & 2000 WK183  &  0.95 &  22.9 &   0.014 &   0.034 \\
134860  & -           & 2000 OJ67   &  0.78 &  18.5 &   0.012 &   0.027 \\
148780  & Altjira     & 2001 UQ18   &  0.90 &  42.4 &   0.344 &   0.042 \\
160091  & -           & 2000 OL67   &  0.76 &  58.0 &   0.240 &   0.044 \\
160256  & -           & 2002 PD149  &  0.83 & 157 &     0.588 &   0.068 \\
275809  & -           & 2001 QY297  &  0.91 &  61.7 &   0.418 &   0.050 \\
364171  & -           & 2006 JZ81   &  0.64 & 330 &     0.850 &   0.043 \\
385446  & Manw\"{e}       & 2003 QW111  &  0.59 &  59.1 &   0.563 &   0.025 \\
469514  & -           & 2003 QA91   &  0.96 &   8.6 &   0.020 &   0.021 \\
469705  & \kagara     & 2005 EF298  &  0.76 &  63.4 &   0.690 &   0.034 \\
508788  & -           & 2000 CQ114  &  0.95 &  86.8 &   0.095 &   0.064 \\
508869  & -           & 2002 VT130  &  0.82 &  13.3 &   0.019 &   0.024 \\ 
524366  & -           & 2001 XR254  &  0.82 &  59.9 &   0.556 &   0.042 \\
524531  & -           & 2002 XH91   &  0.62 &  92.8 &   0.710 &   0.029 \\
525462  & -           & 2005 EO304  &  0.51 & 599 &     0.210 &   0.067 \\
-       & -           & 1998 WW31   &  0.83 & 167 &     0.819 &   0.049 \\
-       & -           & 1999 RT214  &  0.69 &  40.2 &   0.300 &   0.032 \\
-       & -           & 2000 CF105  &  0.79 &  604 &    0.330 &   0.145 \\
-       & -           & 2000 QL251  &  0.97 &  34.2 &   0.489 &   0.035 \\
-       & -           & 2001 QW322  &  0.98 &  804 &    0.464 &   0.179 \\
-       & -           & 2003 QY90   &  0.99 & 106 &     0.663 &   0.056 \\
-       & -           & 2003 TJ58   &  0.79 &  66.0 &   0.516 &   0.043 \\
-       & -           & 2003 UN284  &  0.67 & 521 &     0.380 &   0.107 \\
-       & -           & 2006 BR284  &  0.79 & 316 &     0.275 &   0.107 \\
-       & -           & 2006 CH69   &  0.82 & 297 &     0.896 &   0.049 \\
\hline \hline
\end{tabular}
}
\caption{The equal-size binaries in the CC population of the Kuiper belt. The binary semimajor axis 
$a_{\rm b}$ is scaled by $R_{\rm b}=R_1+R_2$, where $R_1$ and $R_2$ are the estimated radii of binary components. 
The eccentricity of binary orbits, $e_{\rm b}$, formally given to three decimal digits in column 6, is often 
uncertain. In the last column, $l_{\rm b}=L_{\rm b}/L_{\rm H}$ is the scaled angular momentum, where
$L_{\rm b}=\mu_{\rm b} [G M_{\rm b} a_{\rm b} (1-e_{\rm b}^2) ]^{1/2}$ is the angular momentum of the binary orbit, 
$\mu_{\rm b}=M_1 M_2/M_{\rm b}$, $M_{\rm b}=M_1+M_2$, $M_1$ and $M_2$ are the masses of binary components,
and $L_{\rm H}=M_{\rm b} R_{\rm H}^2 \Omega$ (Sect.\ 6).}
\end{table} 

\begin{table}
\centering
{
\begin{tabular}{lrrrrrrrrr}
\hline \hline
run          & domain size              &  number of               & number of        & $\tau$ & $Z$  & $\Pi$ & $\tilde{G}$ & $t_0$   \\   
& ($H_{\rm g}$)                    &  grid cells              & particles        &        &      &       &                          &  ($\Omega^{-1}$)  \\  
\hline
A12          & $0.1\times0.1\times0.2$ & $512\times512\times1024$  & $1.34\times10^8$  &  2    & 0.1    & 0.05 & 0.05       & 36  \\
B22          & $(0.2)^3$                & $(512)^3$                 & $1.54\times10^8$  &  2    &  0.1  & 0.05 & 0.05       & 37  \\
C203         & $(0.2)^3$                & $(512)^3$                 & $1.54\times10^8$  &  0.3  & 0.02   & 0.05 & 0.05      & 110 \\
\hline \hline
\end{tabular}
}
\caption{Model parameters of the three SI simulations used in this work. The domain size is given in the units of gas scale height, 
$H_{\rm g}$. The particle concentration $Z = \Sigma_{\rm p}/\Sigma_{\rm g}$ is the ratio of the particle mass surface density $\Sigma_{\rm p}$ 
to the gas surface density $\Sigma_{\rm g}$. The parameter $\Pi$ expresses the degree to which the gas orbital velocity is 
sub-Keplerian due to the gas pressure gradient. The parameter $\tilde{G}$ quantifies the relative strength of self-gravity to tidal 
shear. It is related to the Toomre $Q$ via $\tilde{G}=(4/\sqrt{2 \pi})Q^{-1}$. The particle gravity was switched on at time $t_0$.} 
\end{table}

\begin{table}
\centering
{
\begin{tabular}{lrrrrrrrrr}
\hline \hline
id & $l_0$ & $l_{0.5}$ & $M/M_{\rm local}$ & $N_{\tt ATHENA}$ & $N_{\rm cl}$  & $R_2/R_1$ & $l_{\rm b}$ & $l_{\rm b}/l_0$ & $M_{\rm b}/M_0$  \\  
   &       &          & $\times 10^{-7}$       &                &             &           &       &          &            \\ 
\hline
\multicolumn{10}{c}{\it Run A12 with $t_{A2P}=40/\Omega$}   \\
a00 & 0.102 & 0.102 & 4.26  & 22788 & 10                           & 0.706     & 0.059 &  0.58    & 0.98       \\
a01 & 0.256 & 0.096 & 3.99  & 21363 & 10                           & 0.376     & 0.017 &  0.07    & 0.76       \\
a02 & 0.037 & 0.035 & 4.53  & 24250 & 10                           & -         & -     & -        & -          \\
a03 & 0.148 & 0.114 & 4.50  & 24073 & 10                           & 0.717     & 0.065 &  0.44    & 0.78       \\
a04 & 0.067 & 0.025 & 3.88  & 20779 & 10                           & -         & -     & -        & -          \\
a05 & 0.103 & 0.031 & 1.58  & 8452  & 30                           & -         & -     & -        & -          \\
a06 & 0.106 & 0.100 & 65.1  & 348545 & 1                           & 0.561     & 0.049 &  0.46    & 0.93       \\ 
a07 & 0.111 & 0.027 & 0.29  & 1540  & 100                          & -         & -     & -        & -          \\
a08 & 0.107 & 0.107 & 12.2  & 65107 &  3                           & 0.741     & 0.078 &  0.73    & 0.96       \\
a09 & 0.106 & 0.061 & 3.66  & 19577 & 10                           & -         & -     & -        & -          \\
a10 & 0.097 & 0.097 & 4.81  & 25763 & 10                           & 0.761     & 0.061 &  0.63    & 0.93       \\   
\multicolumn{10}{c}{\it Run A12 with $t_{A2P}=38/\Omega$}   \\
a00 & 0.314 & 0.110 & 5.06  & 27103  & 10                          & 0.834     & 0.069 & 0.22     & 0.63       \\
a01 & -     & -     & -     &     -  & -                           & -         & -     & -        & -          \\
a02 & 0.152 & 0.075 & 9.42  & 50460  &  5                          & 0.690     & 0.056 & 0.37     & 0.73       \\
a03 & 0.275 & 0.097 & 4.23  & 22672  & 10                          & 0.694     & 0.045 & 0.16     & 0.59       \\
a04 & 0.436 & 0.046 & 24.0  & 128313 &  2                          & -         & -     & -        & -          \\
a05 & 0.408 & 0.081 & 1.81  & 9705   & 25                          & 0.713     & 0.116 & 0.28     & 0.58       \\
a06 & 0.080 & 0.064 & 37.9  & 202799 & 1                           & 0.737     & 0.079 & -        & 0.88       \\
a07 & 0.104 & 0.040 & 13.9  & 74386  &  3                          & 0.366     & 0.027 & 0.26     & 0.87       \\
a08 & 0.175 & 0.131 & 12.1  & 64593  &  3                          & 0.728     & 0.079 & 0.45     & 0.74       \\
a09 & 0.180 & 0.037 & 3.49  & 18701  & 10                          & -         & -     & -        & -          \\
a10 & 0.180 & 0.097 & 4.70  & 25190  & 10                          & 0.587     & 0.032 & 0.18     & 0.74       \\   
\hline \hline
\end{tabular}
}
\caption{The pebble clumps selected for the {\tt PKDGRAV} simulations: run A12. In the third column, $l_{0.5}$ denotes the SAM content
in a sphere with radius $r=0.5 R_{\rm H}$ (Section 8.6). In the fourth column, $N_{\tt ATHENA}$ denotes the number 
of {\tt ATHENA} particles in each clump. The cloning factor, $N_{\rm cl}$, is given in the fifth column. The binary properties 
listed in the last four columns were obtained by taking the mean over the results of our five simulations with different seeds.
In the last column, $M_{\rm b}/M_0$ is the ratio of the final binary and initial clump masses. The final binary mass represents 
58-98\% of the initial clump mass. The masses of binaries reported here are between $5 \times 10^{20}$ g and $1.5 \times 10^{22}$ 
g for the disk model discussed in Sect.\ 7. For comparison, the least and most massive binaries in Table 1 -- 2000 CF105 and 
(79360) Sila-Nunam -- have masses $2.2 \times 10^{21}$ g and $1.1 \times 10^{22}$ g, respectively.} 
\end{table}

\begin{table}
\centering
{
\begin{tabular}{lrrrrrrrrr}
\hline \hline
id & $l_0$ & $l_{0.5}$ & $M/M_{\rm local}$ & $N_{\tt ATHENA}$ & $N_{\rm cl}$  & $R_2/R_1$ & $l_{\rm b}$ & $l_{\rm b}/l_0$ & $M_{\rm b}/M_0$  \\  
   &       &          & $\times 10^{-7}$ &               &             &           &            &               &                \\ 
\hline
\multicolumn{10}{c}{\it Run B22 with $t_{A2P}=41/\Omega$}   \\
b00 & 0.117 & 0.083   & 20.8            & 27809         &  11          & 0.753    & 0.048      & 0.41          & 0.84           \\
b01 & 0.272 & 0.129   & 28.0            & 37432         &  8           & 0.721    & 0.050      & 0.19          & 0.75           \\
b02 & 0.116 & 0.107   & 6.24            & 8351          &  36          & 0.987    & 0.092      & 0.80          & 0.74           \\
b03 & 0.212 & 0.097   & 4.05            & 5427          &  55          & 0.355    & 0.008      & 0.04          & 0.74           \\
b04 & 0.071 & 0.056   & 7.03            & 9406          &  32          & 0.673    & 0.040      & 0.57          & 0.82           \\
b05 & 0.082 & 0.076   & 26.4            & 35296         &  8           & 0.506    & 0.042      & 0.51          & 0.96           \\
b06 & 0.080 & 0.080   & 196.            & 262936        &  1           & 0.884    & 0.060      & 0.75          & 0.97           \\
b07 & 0.090 & 0.075   & 101.            & 135266        &  2           & 0.576    & 0.035      & 0.39          & 0.92           \\
b08 & 0.059 & 0.036   & 38.7            & 51788         &  6           & 0.202    & 0.006      & 0.11          & 0.29           \\
b09 & 0.202 & 0.197   & 97.5            & 130136        &  2           & 0.816    & 0.079      & 0.39          & 0.75           \\  
b10 & 0.114 & 0.043   & 11.3            & 15089         &  20          & 0.365    & 0.005      & 0.04          & 0.82           \\
b11 & 0.180 & 0.055   & 4.22            & 5647          &  53          & 0.477    & 0.023      & 0.13          & 0.78           \\
b12 & 0.034 & 0.031   & 92.3            & 123577        &  2           & -        & -          & -             & -              \\
b13 & 0.163 & 0.083   & 10.6            & 14130         &  21          & 0.935    & 0.063      & 0.39          & 0.72           \\
b14 & 0.108 & 0.061   & 11.5            & 15442         &  19          & 0.459    & 0.015      & 0.14          & 0.85           \\ 
b15 & 0.170 & 0.130   & 113             & 150903        &  2           & 0.957    & 0.141      & 0.83          & 0.93           \\ 
\multicolumn{10}{c}{\it Run C203 with $t_{A2P}=117.6/\Omega$}   \\
c00 & 0.083 & 0.083   & 23.6            & 157976        &  2           & 0.548    & 0.027      & 0.33          & 0.91           \\ 
c01 & 0.171 & 0.144   & 23.5            & 157869        &  2           & 0.665    & 0.146      & 0.85          & 0.50           \\
c02 & 0.173 & 0.148   & 3.27            & 21870         & 14           & 0.703    & 0.069      & 0.40          & 0.57           \\
c03 & 0.103 & 0.103   & 51.1            & 341809        &  1           & 0.667    & 0.051      & 0.50          & 0.91           \\
c04 & 0.102 & 0.065   & 6.94            & 46430         &  6           & 0.345    & 0.009      & 0.09          & 0.85           \\
c05 & 0.057 & 0.057   & 45.1            & 301838        &  1           & 0.588    & 0.039      & 0.68          & 0.97           \\
c06 & 0.163 & 0.108   & 31.1            & 208462        &  1           & 0.706    & 0.060      & 0.37          & 0.85           \\
c07 & 0.058 & 0.041   & 4.57            & 30602         & 10           & 0.364    & 0.011      & 0.20          & 0.727          \\  
\hline \hline
\end{tabular}
}
\caption{The pebble clumps selected for the {\tt PKDGRAV} simulations: runs B22 and C203. See caption of Table 3 for additional
information. Unlike in Table 3 the last four columns report results for individual {\tt PKDGRAV} simulations with single
seeds (no averaging); larger variability is expected here.} 
\end{table}

\clearpage

\begin{figure}
\epsscale{0.48}
\plotone{fig11a.eps}
\plotone{fig11b.eps}
\caption{The SAM distribution of clumps identified for $t_{\rm A2P}=38/\Omega$ (black dots; panel a) and $t_{\rm A2P}=40/\Omega$
(panel b). The SAM content within a sphere of radius $r=0.5 R_{\rm H}$ is shown on the $y$ axis ($l_{0.5}$). The green 
stars label the clumps that produced equal-size binaries with $R_2/R_1>0.5$, the blue stars label unequal-size
binaries and hierarchical systems with $R_2/R_1<0.5$, and the red stars label the clumps that produced single 
planetesimals ($R_2<10$ km; nominal {\tt PKDGRAV} resolution used here). The dashed lines approximately divide 
parameter space into domains where equal-size binaries do ($l_{0.5}>0.07$) and do not form ($l_{0.5}<0.07$). The concentration 
of points along the linear feature in (b) happens because many clumps are already compact at $t_{\rm A2P}=40/\Omega$, and 
$l_{0.5} \simeq l$.}
\label{criter}
\end{figure}

\begin{figure}
\epsscale{0.48}
\plotone{fig16a.eps}
\plotone{fig16b.eps}
\caption{The SAM distribution of B22 clumps identified for $t_{\rm A2P}=41/\Omega$ (panel a) and C203 clumps 
for $t_{\rm A2P}=117.6/\Omega$ (panel b). The SAM content within a sphere of radius $r=0.5 R_{\rm H}$ in shown on the 
$y$ axis ($l_{0.5}$). The green stars label the clumps that produced the equal-size binaries with $R_2/R_1>0.5$, the blue 
stars label the unequal-size binaries and hierarchical systems with $R_2/R_1<0.5$, and the red stars label the clumps that 
produced single planetesimals ($R_2<10$ km; nominal {\tt PKDGRAV} resolution used here). The dashed lines 
approximately divide parameter space into domains where equal-size binaries do ($l_{0.5}>0.07$) and do not 
form ($l_{0.5}<0.07$).}
\label{criter2}
\end{figure}

\end{document}